\pdfoutput=1
\documentclass[11pt]{arxiv-article}
	
\bibliography{References}

\usepackage{Definitions}
\usepackage{tikz}
\usetikzlibrary{calc}
\usepackage{pifont}
\usepackage{hyperref}
\usepackage{mathrsfs}  

\allowdisplaybreaks

\newcommand{\OfficialTitle}{
 Resurgence of the large-charge expansion
}
\title{\setstretch{1.4}
  {\color{Thoughtless}\Huge\textbf{\dosserif\OfficialTitle}}
}

\hypersetup{pdfauthor={},pdftitle={\OfficialTitle}}

\author{%
  \begin{minipage}{.97\linewidth}
    \vspace{1cm}
    \begin{center} \dosserif%
      {\small
       \textbf{Nicola~Dondi}\textsuperscript{\ding{73}},
        \textbf{Ioannis~Kalogerakis}\textsuperscript{\ding{73}},
        \textbf{Domenico~Orlando}\textsuperscript{\ding{72}\ding{73}},
         and
         \textbf{Susanne~Reffert}\textsuperscript{\ding{73}} 
         }
    \end{center}
    \vspace{1cm}
        \authorBlock{\ding{73}}{\dosserif{} Albert Einstein Center for Fundamental Physics\\
      Institute for Theoretical Physics, University of Bern,\\
      Sidlerstrasse 5, CH-3012 Bern, Switzerland}
    \authorBlock{\ding{72}}{\dosserif{} INFN sezione di Torino | Arnold--Regge Center\\
      via Pietro Giuria 1, 10125 Torino, Italy}
  \end{minipage}
}

\date{}

\begin{document}

\setstretch{1.2}

\numberwithin{equation}{section}

\begin{titlepage}

  \newgeometry{top=23.1mm,bottom=46.1mm,left=34.6mm,right=34.6mm}

  \maketitle

  \thispagestyle{empty}

  \vfill\dosserif{}

  \abstract{\normalfont{}\noindent{}%
    We study the \(O(2N)\) model at criticality in three dimensions in the double scaling limit of large \(N\) and large charge. We show that the large-charge expansion is an asymptotic series, and we use resurgence techniques to study the non-perturbative corrections and to extend the validity of the \acs{eft} to any value of the charge. We conjecture the general form of the non-perturbative behavior of the conformal dimensions for any value of \(N\) and find very good agreement with previous lattice data. }

\vfill

\end{titlepage}

\restoregeometry{}

\setstretch{1.2}

\tableofcontents

\section{Introduction}%
\label{sec:introduction}

Working in sectors of large global charge leads to important simplifications in strongly coupled and otherwise inaccessible \acp{cft}~\cite{Hellerman:2015nra,Alvarez-Gaume:2016vff,Gaume:2020bmp}.

One striking feature of the large-charge expansion of the O(2N) model at the \ac{wf} fixed point in 2+1 dimensions is that it appears to also work for small charges, which taken as such is quite astounding: in general we expect the semiclassical expansion to work for systems with a very large number of degrees of freedom.
This fact came to light first in the comparison of the large-charge prediction for the scaling dimension of the lowest operator of charge $Q$ with lattice results for the O(2)~\cite{Banerjee:2017fcx} and the O(4)~\cite{Banerjee:2019jpw} models.
It was noted that very few terms in the effective action were sufficient to reproduce the lattice results with very high accuracy.

From the \ac{eft} point of view, there seems to no reason for the large-charge predictions to keep working for small charges.
But if we add another controlling parameter to the mix, we can go beyond the reach of the \ac{eft} and try to understand this behavior.
This is for example the case when we study the large-$N$ limit of the O(2N) model at large charge~\cite{Alvarez-Gaume:2019biu,Giombi:2020enj}.
In the double-scaling limit $Q\to \infty$, $N\to \infty$, with $Q/(2N) = \hat q $ constant it is possible to solve the problem exactly at leading order in \(N\) for any value of \(\hat q\).
Building on these results, in this paper we show that the large-charge expansion in the double-scaling limit is asymptotic, and is closely related to the asymptotic Seeley--DeWitt expansion of heat kernels and related $\zeta$-functions on spheres~\cite{CANDELAS1984397,Dowker:2003ra}. 

  Asymptotic series are a common feature of perturbative solutions to quantum problems as originally argued by Dyson~\cite{Dyson:1952tj}. This feature signals the presence of non-perturbative phenomena in the theory, as was first quantitatively discussed in a series of papers by Bender and Wu starting with~\cite{Bender:1971gu} in the context of anaharmonic oscillators.
The modern approach to the subject goes under the name of \emph{resurgent asymptotics}, or simply resurgence, and originates from Écalle's works~\cite{Ecalle80}.
Modern reviews on the subject and on applications in physics and mathematics are~\cite{Dorigoni:2014hea,Aniceto:2018bis} to which we refer for a complete list of references.
We will use resurgent methods to show which non-perturbative contributions are present in the double-scaling limit of the large-charge expansion and how non-perturbative ambiguities cancel.
The final result is an extrapolation to small charge which matches the small-charge expansion with excellent precision.

We develop a geometric picture interpreting the exponential corrections to the perturbative series as worldline instantons describing a particle of mass equal to the chemical potential $\mu$ moving along geodesics.
We also find a geometric interpretation of the lateral Borel resummation in terms of unstable worldline  modes.

We conjecture this picture to be robust enough to carry over to the general case of finite \( N\), lending credibility to the \emph{general validity} of our observations made in the double-scaling limit.
Assuming that the qualitative features of the worldline instanton persist also for finite \(N\), we can use the knowledge of the leading exponential effects to derive constraints on the perturbative expansion, even though in this case, the coefficients are not accessible within the \ac{eft}.
We find that a few terms (of order \(N^* \approx \sqrt{Q}\)) are enough to describe the theory also at small charge with a precision of order \(1\%\), in agreement with the lattice estimates.
The same geometric picture also appears in the case of supersymmetric systems.
For example, in \(\mathcal{N}=2\) \textsc{sqcd} at large R-charge similar non-perturbative contributions were discussed numerically~\cite{Hellerman:2017sur,Hellerman:2018xpi,Hellerman:2020sqj} and analytically in the double scaling limit in~\cite{Bourget:2018obm,Grassi:2019txd,Beccaria:2018xxl,Beccaria:2020azj} and were found to be consistent with the propagation of massive hypermultiplets around geodesics.

	It is interesting that the leading non-perturbative effects are ``classical'', in the sense that they are a consequence of the effective action itself being an asymptotic expansion, growing as $(2n)!$.
	They scale as $e^{-\hat q^{1/2}}$, while more traditional instantons associated to Feynamn diagram proliferation are controlled by \(e^{-\hat q^{3/2}}\) and are thus suppressed.
	The same $(2n)!$ factorial growth is found also in computations of effective actions of the Euler--Heisenberg type~\cite{Heisenberg:1935qt}, see also \cite{Dunne:2004nc} for a comprehensive review. In those instances, it has been shown \cite{Dunne:2005sx,Dunne:2006st} that this faster-than-factorial growth is driven by worldline instantons, as it turns out to be also in the present work. 
	This is also reminiscent of the situation in string theory, where D-instantons generically dominate over NS5 and gravitational instantons.%

\bigskip
The plan of this note is as follows.
In Section~\refstring{sec:asymptotics}, we study the asymptotics of the O(2N) model in the double-scaling limit with resurgence methods.
As specific examples, we discuss the model on the torus (Section~\ref{sec:torus}) and on the sphere (Section~\ref{sec:perturbative-sphere}), giving both the perturbation series and the exponentially suppressed non-perturbative corrections.
These examples have special features due to their geometry --- the perturbation series of the torus consists of a single term due to the vanishing of the curvature. The case of the sphere is more involved --- in fact, we need to supplement our resurgence calculation with a physical input on how to resolve the ambiguities. 
This input is provided in Section~\refstring{sec:worldline}, where we reformulate our problem as the quantum mechanics of a particle moving on closed geodesics. We recast the heat-kernel as a path integral (Section~\ref{sec:path-integral}) and discuss again the torus (Section~\ref{sec:torus-path-integral}) and sphere (Section~\ref{sec:sphere-path-integral}) examples. 
In Section~\refstring{sec:resurgence}, we combine the results of the previous two sections and obtain the exact form of the grand potential for any value of the charge. This is substantiated numerically against the small-charge expansion. 
In Section~\refstring{sec:lessons-from-large-N}, we go beyond the double-scaling limit by arguing that the geometric interpretation in terms of worldline geodesics works for the general case. Assuming the large-charge expansion to always be an asymptotic series for any $N$, we can give the optimal truncation as a function of the charge. This result is borne out by the observations made on the lattice.
In Section~\refstring{sec:conclusions}, we summarize our result and point out future lines of investigations.

In the appendix we summarize the prerequisites of large charge at large N (Appendix~\ref{sec:largeN}), introduce some basic facts about resurgence (Appendix~\ref{sec:borel-transform}), discuss Lipatov's instantons (Appendix~\ref{sec:lipatovs-instantons}), discuss the optimal truncation in the double-scaling limit (Appendix~\ref{sec:optimal-truncation}), and present a simple worked-out example of resurgent analysis which is relevant for Section \refstring{sec:asymptotics}  (Appendix~\ref{sec:Dawson}).

\newpage

\section{Asymptotics at large charge}%
\label{sec:asymptotics}

We start with the Landau--Ginzburg model for \(2N\) real scalar fields in the vector representation of \(O(2N)\) (which we encode as \(N\) complex fields) in \((1+2) \) dimensions with Euclidean signature on $(\setR \times \mani)$, where $\mani$ is a Riemann surface.
Including all terms up to mass dimension three, we have
\begin{equation}\label{eq:UV-Hamiltonian}
  S[\varphi_i] = \sum_{i=1}^N  \int \dd{t} \dd{\mani} \bqty{g^{\mu\nu} \pqty{\del_\mu \varphi_i}^* \pqty{\del_\nu \varphi_i} + r \varphi_i^* \varphi_i  + \frac{u}{2}  \pqty{\varphi_i^* \varphi_i}^2 + \frac{v}{4}  \pqty{\varphi_i^* \varphi_i}^3}.
\end{equation}
This model flows to the \ac{wf} fixed point in the \ac{ir} limit \(u \to \infty\) when \(r\) is fine-tuned to the value of the conformal coupling.
\subsection{Results from large N}
\label{sec:results-from-large-N}

We are interested in the free energy of the O(2N) vector model at criticality.
We want to work in a sector of fixed charge \(Q\) which corresponds to considering the completely symmetric representation of rank $Q$ of the global symmetry group O(2N)~\cite{Banerjee:2019jpw,Alvarez-Gaume:2019biu,Antipin:2020abu,Giombi:2020enj}. We take the double-scaling limit 
\begin{align}
	N &\to \infty,  & Q &\to \infty, & \hat q &= Q/(2N) \ \text{fixed}.
	\label{eq:limit}
\end{align}
The main result of~\cite{Alvarez-Gaume:2019biu} is that the free energy can be expressed as the Legendre transform of a zeta function (see Appendix~\ref{sec:largeN} for details).
Let \(f(\hat q) = F/(2N)\) be the free energy per \ac{dof}, \(\omega(\mu)\) the grand potential, and \(\mu\) the chemical potential. We have
\begin{align}
  f(\hat q ) &= \sup_\mu (\mu \hat q - \omega(\mu)), \label{eq:hatq}\\
  \hat q &= \dv{\omega(\mu)}{\mu}, \\
  \omega(\mu) &= -\frac{1}{2} \zeta(- \tfrac{1}{2} | \mani, \mu) , \label{eq:grandPot}
\end{align}
where \(\zeta(s| \mani, \mu)\) is the zeta function for the operator \(-\Laplacian{} + \mu^2\), \emph{i.e.} if \(\spec(- \Laplacian{}) = \set{\lambda_j}\),
\begin{equation}
  \zeta(s | \mani, \mu)   = \sum_{j} (\lambda_j + \mu^2)^{-s}.
\end{equation}
The large-\(\hat q\) regime corresponds to large chemical potential \(\mu\).
In this case it is convenient to write the zeta function in the Mellin representation
\begin{equation}
  \zeta(s| \mani, \mu) = \frac{1}{\Gamma(s)} \int_0^\infty \frac{\dd{t}}{t} t^s e^{-\mu^2 t}  \Tr(e^{\Laplacian{} t}) .
\end{equation}
If \(\mu\) is large, the integral localizes around \(t = 0\).
This reduces the large-charge problem to the classical problem of the Weyl asymptotic expansion of the heat kernel, which can be written in terms of Seeley--DeWitt coefficients~\cite{DeWitt:1965jb,Seeley:1967ea}:
\begin{equation}
  \Tr(e^{\Laplacian{} t}) \sim \frac{V}{4 \pi t} \pqty{ 1 + \frac{R}{12} t + \dots } .
\end{equation}
In the following we will concentrate on the examples of the torus \(\mani = T^2\) and the sphere \(\mani = S^2\).
We choose the former because of its simplicity and the latter because we can identify the free energy on the sphere with the conformal dimension of the lowest operator of charge \(Q\) using the state-operator correspondence.
In the case of \(T^2\), the Weyl expansion contains only the leading term because all the others are proportional to the curvature.
In the case of \(S^2\) we will see that the series is asymptotic and can be studied with the methods of resurgence theory.

\subsection{The torus}
\label{sec:torus}

Our first example is the torus \(\mani = T^2\).
This case is particularly simple because the free energy can be written exactly, but it is interesting because it shows some qualitative properties that are general.

Consider a square torus of side \(L\). All but the first Seeley--DeWitt coefficients vanish, so the leading asymptotic of the heat trace and $\zeta$-function are
\begin{align}
  \Tr(e^{\Laplacian{} t}) &\sim \frac{L^2}{4 \pi t}  + \order{e^{-1/t}}, &   \zeta(s| T^2, \mu) &=  \frac{L^2 \mu^{2(1-s)}}{4 \pi(s-1)}  + \order{e^{-\mu}}.
 \label{eq:LOT2}
\end{align}
One can readily derive all the quantities appearing in the system Eq.~\eqref{eq:hatq}-\eqref{eq:grandPot}:
\begin{align}
  f(\hat q) &= \frac{4 \sqrt{\pi}}{3L} \hat q^{3/2},\\
  \hat q &= \frac{L^2 \mu^2}{4 \pi} ,\\
   \omega(\mu) &= - \frac{1}{2} \zeta(-\tfrac{1}{2}| T^2, \mu)  = \frac{L^2 \mu^3}{12 \pi} .
\end{align}
These expressions are perturbatively exact at leading order in large $\mu$ (respectively $\hat{q}$), but one can do much better since the $\order{e^{-1/t}}$ corrections in Eq.~\eqref{eq:LOT2} are known in closed form.
From the spectrum of the $T^2$ Laplacian,
\begin{equation}
	\spec(\Laplacian_{T^2}) = \left\{ - \frac{4 \pi^2}{L^2} (k_1^2 + k_2^2) \middle| k_1, k_2 \in \setZ \right\},
\end{equation}
follows that the heat kernel trace is the square of a theta function:
\begin{equation}
  \Tr(e^{\Laplacian{} t })  = \sum_{k_1,k_2 \in \setZ} e^{-\frac{4 \pi^2}{L^2} (k_1^2 + k_2^2) t } = \bqty{ \theta_3(0, e^{-\frac{4 \pi^2 t}{L^2}})}^2.
\end{equation}
After Poisson resummation, we find the appropriate expansion around $t \rightarrow 0^+$, 
\begin{equation}
  \label{eq:Torus-trace-Poisson}
  \Tr(e^{\Laplacian{} t })  = \frac{L^2}{4 \pi t} \pqty{1 + \sideset{}{'}\sum_{\mathbf{k} \in \setZ^2} e^{- \frac{\norm{\mathbf{k}}^2 L^2}{4 t} }},
\end{equation}
where $\norm{ \mathbf{k}}^2 = k_1^2 + k_2^2$, and the prime indicates that the sum does not include the zero mode.
This expression is exact and valid also for finite $t$, which allows us to find the subleading contribution for $\mu \rightarrow \infty$ of the $\zeta$-function:
\begin{equation}
  \zeta(s| T^2, \mu) = \frac{L^2 \mu^{2 - 2s}}{4 \pi ( s -1)} + \frac{L^2}{2 \pi} \sideset{}{'}\sum_{\mathbf{k}} \frac{2^{2-s}}{\Gamma(s)} \pqty{\frac{\norm{\mathbf{k}}L}{\mu} }^{s-1} K_{1-s}(\norm{\mathbf{k}} \mu L) ,
\end{equation}
where \(K_n(z)\) is the modified Bessel function of the second kind. The subleading terms in the grand potential and the free energy are then given in closed form by
\begin{align}
  \omega(\mu) &= - \frac{1}{2}  \zeta(-\tfrac{1}{2} | T^2, \mu) = \frac{L^2 \mu^3}{12 \pi} \pqty{ 1 + \sideset{}{'}\sum_{\mathbf{k}} \frac{e^{-\norm{\mathbf{k}} \mu L }}{\norm{\mathbf{k}}^2 \mu^2 L^2 } \pqty{1 + \frac{1}{\norm{\mathbf{k}} \mu L} }  }, \\
  f(\hat q) &= \sup_\mu( \mu \hat q - \omega(\mu)) = \frac{4 \sqrt{\pi}}{3 L} \hat q^{3/2} \pqty{1 - \sideset{}{'}\sum_{\mathbf{k}} \frac{e^{- \norm{\mathbf{k}} \sqrt{4 \pi \hat q}}}{8 \norm{\mathbf{k}}^2 \pi \hat q} + \dots  } .
\end{align}
The relation between the Legendre-conjugate variables $\hat{q}$ and $\mu$ is computed recursively, so that further corrections to the free energy are present. However, even extrapolating to small $\hat{q}$ up to \(\hat q = 1\), the contribution of the further exponentially-suppressed terms is of order \(\order{10^{-3}}\).

This is the generic form that we will encounter:
a perturbative expansion in \(\mu\) (which here contains a single term) plus exponentially suppressed terms controlled by the dimensionless parameter \(\mu L\), where \(L\) is the typical scale of the manifold \(\mani\).
Equivalently, the free energy is written as a double expansion in the two parameters \(1/\hat q\) and \(e^{-\sqrt{4 \pi \hat q}}\).\footnote{Note the square root in the exponential, suggesting that there are non-perturbative effects more important than the ``usual'' \ac{qft} instantons that would be generically of order \(\order{e^{- \hat q^{3/2}}}\), see Appendix~\ref{sec:lipatovs-instantons}.}
These structures are known as \emph{trans-series} and appear naturally in perturbative problems. In general, given a problem with a small parameter $z$, a trans-series solution for $z \rightarrow 0$ has the form
\begin{equation}
	\Phi(\sigma_k,z) = \Phi^{(0)}(z) +  \sum_{k \neq 0} \sigma_k e^{-A_k/z^{1/\beta_k}} z^{-b_k/\beta_k} \Phi^{(k)}(z).
	\label{eq:trans-series}
\end{equation} 
Generically, all $\Phi^{(k)}(z)$ are asymptotic series, in particular $\Phi^{(0)}$ is the (formal) solution of the perturbative problem. For this reason, these expressions make sense only once a prescription for the summation of these series is given. We will discuss this issue in the next section.
\emph{Resurgent trans-series} are characterized by a specific set of relations between the numbers $A_k, \beta_k , b_k$ and the series $\Phi^{(k)}$. Thanks to these relations, it is possible to fix the value of the \emph{trans-series parameters} $\sigma_k \in \mathbb{C}$ such that an appropriate summation of the trans-series produces an unambiguous full function which has $\Phi^{(0)}$ as leading perturbative asymptotic expansion. Such a function is known as  \emph{resurgent function}. For a more complete treatment of the subject we refer to~\cite{Dorigoni:2014hea,edgar2009transseries}.

The heat trace on $T^2$ indeed has this form, but trivially all $\Phi^{(k )}$ are one-loop exact.
No ambiguities related to asymptotic series are present, so the resurgent function solution coincides with its trans-series representation, which is just the function $\theta_3$. 
As we will see, this simple result does not carry over to the case of \(\mani = S^2\).

\subsection{The sphere}
\label{sec:perturbative-sphere}

Next we study the case of the sphere of radius \(r\), \(\mani = S^2\).
Now the small-\(t\) behavior of the heat kernel can be represented in terms of an asymptotic series, whose coefficients are the well-known Seeley--DeWitt coefficients for $S^2$.
We will show how the heat trace can be recast into a trans-series form.
This series is not Borel resummable, and we will need to supplement the perturbative expansion with non-perturbative exponentials in order to make sense of it.
These exponentials have a clear interpretation in terms of \acl{wl} instantons, and the non-perturbative ambiguities are related to tachyonic instabilities as we will see in Section~\ref{sec:path-integral}.

\paragraph{Seeley--DeWitt coefficients.}

The spectrum of the Laplacian on the two-sphere is given by
\begin{equation}
	\spec(\Laplacian_{S^2}) = \left\{ - \frac{\ell(\ell +1 )}{r^2} \middle| \ell \in \setN \right\}
\end{equation}
and each eigenvalue has multiplicity \((2 \ell + 1)\).
Like in the case of the torus, the trace sum can be rewritten using Poisson resummation after some massaging (see \emph{e.g.}~\cite{Marklof:2004xxx}).

It is convenient to consider the trace of the conformal Laplacian, which reads
\begin{equation}
	\begin{aligned}
		\Tr \left[ e^{ \left( \Laplacian_{S^2} - \tfrac{1}{4r^2} \right) t} \right] &=  \sum_{\ell = 0}^\infty (2\ell +1)  e^{- \frac{t}{r^2} \left( \ell + \frac{1}{2} \right)^2} = \sum_{\ell = - \infty}^{\infty} \abs{ \ell + \frac{1}{2}} e^{- \frac{t}{r^2} \left( \ell + \frac{1}{2} \right)^2} \\
		&=  \int_\setR \dd{\rho} \abs{\rho} e^{-\rho^2 t/r^2} + \sideset{}{'} \sum_{k \in \setZ} (-1)^k \int_\setR \dd{\rho} \abs{\rho}   e^{-\rho^2 t/r^2 + 2 \pi i k \rho } \\
		&= \frac{r^2}{t} + \sideset{}{'} \sum_{k \in \setZ} (-1)^k \bqty{ \frac{r^2}{t} - \frac{2 \abs{k} \pi r^3}{t^{3/2}} \operatorname{F}\left( \tfrac{\pi r \abs{k}}{\sqrt{t}} \right)  }.
	\end{aligned}
\end{equation}
In the last line we have introduced the Dawson's function\footnote{See Appendix~\ref{sec:Dawson} for a summary of the properties of $F(z)$ and on the construction of its trans-series representation.} \(F(z)\) which is related to the imaginary error function:
\begin{equation}
  F(z) = e^{-z^2} \int_0^z \dd{t} e^{-t^2} = \frac{\sqrt{\pi}}{2} e^{-z^2 } \operatorname{erfi}(z) .
\end{equation}
For small values of its argument, we can use the asymptotic expansion of \(F(z)\),
\begin{equation}
  F(z) \sim \sum_{n=0}^\infty \frac{(2n - 1)!!}{2^{n+1}} \pqty{\frac{1}{z} }^{2n + 1}.
  \label{eq:Dawson}
\end{equation}
After some formal manipulations one obtains the leading asymptotic of the heat trace:
\begin{equation}
  \Tr \left[ e^{\left(\Laplacian - \tfrac{1}{4r^2}\right) t} \right] \sim \frac{r^2}{t} - \sum_{n=1}^\infty \frac{(-1)^n(1-2^{1-2n})}{n! r^{2n-2}} B_{2n} t^{n-1} \equiv \frac{r^2}{t}  \sum_{n=0}^\infty a_n \pqty{\frac{t}{r^2}}^n,
  \label{eq:Phi0}
\end{equation}
where \(B_{2n}\) are the Bernoulli numbers.
This expression was already discussed in~\cite{Cahn1975}, based on previous work in~\cite{mulholland1928asymptotic}.
The series is asymptotic since at large $n$ the Seeley--DeWitt coefficients diverge like \(n!\):
\begin{align}
	B_{2n} &= (-1)^{n+1} \frac{2 (2n)!}{(2\pi)^{2n}} \zeta(2n) & \implies &&   a_n &= \frac{(-1)^{n+1}(1-2^{1-2n})}{n!} B_{2n} \sim \frac{2}{ \sqrt{\pi} } \frac{n^{-1/2}}{\pi^{2n} } n!.
	\label{eq:S2coeff}
\end{align}
This divergence is a direct consequence of the expansion of the Dawson's function Eq.~\eqref{eq:Dawson}, which is itself asymptotic.

This expansion is only valid formally and needs a summing prescription.
We assume that this series can be completed into a resurgent trans-series, and that an appropriate summation procedure leads to an unambiguous solution in terms of a resurgent function. 
The first step is to identify the correct form of the non-perturbative corrections in the generic trans-series of~Eq.\eqref{eq:trans-series}, where the perturbative expansion in Eq.~\eqref{eq:Phi0} plays the role of $\Phi^{(0)}$. We will take them to have the general form
\begin{align}
&\sum_{k \neq 0} e^{-A_k/z^{1/\beta_k}} z^{-b_k/\beta_k} \Phi^{(k)}(z) , & \Phi^{(k)}(z) &\sim \sum_{\ell = 0}^\infty a^{(k)}_\ell z^{\ell/\beta_k},
\end{align}
where we employed $z = t / r^2$.
One of the main results concerning resurgent functions is that the coefficients \(a^{(k)}_\ell\) of the series $\Phi^{(k>0)}$ together with the numbers $\beta ,A_k , b_k$ are encoded in the large-order behavior of the perturbative series~\cite{Dorigoni:2014hea}:
\begin{equation}
	a_n \sim \sum_k \frac{S_k}{2 \pi i} \frac{\beta_k}{A_k^{n \beta_k + b_k}} \sum_{\ell = 0}^\infty a_\ell^{(k)} A_k^{\ell} \Gamma( \beta_k n + b_k - \ell),
	\label{eq:GenericLargeOrder}
\end{equation}
where the \(S_k\) are Stokes constants.\footnote{The large-order relation is trivially realized in the case of $T^2$.}
In our case we have complete knowledge of the \(a_n\). Upon using the identity 
\begin{equation}
	\sum_{k\neq 0} \frac{(-1)^k}{k^{2n}} = (2^{-2n+1}-1) (-1)^{n+1} \frac{(2 \pi)^{2n}}{(2n)!} B_{2n},
\end{equation}
we can write them in the suggestive form
\begin{equation}
	a_n = -\frac{1}{\sqrt{\pi}}  \sum_{k\neq0} (-1)^k \frac{\Gamma(n+ \tfrac{1}{2})}{(\pi k)^{2n}}.
\end{equation}
Comparing the two expressions we find
\begin{align}
	\beta &= 1, & b_k&= \frac{1}{2}, & A_k &= (\pi k)^2, & \frac{S_k}{2\pi i } a_0^{(k)} &= (-1)^{k+1}  \abs{k} \sqrt{\pi} , & a^{(k)}_{>0} &= 0.
\end{align}
The series around each exponential are truncated to only one term. This shows that a trans-series representation of the heat trace has to contain the terms
\begin{equation}
	\label{eq:non-perturbative-heat}
 \Tr \left[ e^{\left(\Laplacian - \tfrac{1}{4r^2}\right) t} \right] \,\supset\, 	 2 i\left( \frac{\pi r^2}{t} \right)^\frac{3}{2}  (-1)^{k+1} \abs{k} e^{- (k\pi r)^2/t}.
\end{equation}
These contributions to the trans-series representations are defined up to a $k$-dependent complex constant (the trans-series parameters $\sigma_k$ in Eq.~\eqref{eq:trans-series}).
The large-order analysis of $\Phi^{(0)}$ cannot fix these constants, which is a reflection of the fact that for any choice of $\sigma_k$, the resulting trans-series has $\Phi^{(0)}$ as perturbative asymptotics.

\paragraph{Grand potential and free energy.}

The grand potential and the free energy are themselves asymptotic series. Being related to the Mellin transform of the heat trace, these quantities are higher factorially divergent.
This seems to be a  feature of the model in the double scaling limit Eq.~\eqref{eq:limit}: \emph{the large-charge expansion of the conformal dimension is asymptotic and the coefficients in the series diverge like \((2n)!\)}. 
We will argue in Section~\ref{sec:lessons-from-large-N} that this feature is a general feature of large-charge limits.

A similar large-order analysis can be carried out for the large-\(\mu\) expansion of the zeta function,
\begin{equation}
	\zeta(s | S^2 , \mu) = \frac{1}{\Gamma(s)}\int_0^\infty \dd t \, t^{s-1} e^{-\mu^2 t} \Tr \left[ e^{\Laplacian{} t} \right].
\end{equation}
It is convenient to use the conformal Laplacian, which amounts to a shift $\mu^2 \rightarrow \mu^2 - 1/(4r^2) \equiv m^2$.
One obtains~\cite{Alvarez-Gaume:2019biu}
\begin{equation}  
    \zeta(s | S^2, m)
    = r^2 m^{2(1-s)} \sum_{n=0}^\infty a_n \frac{\Gamma( n+s-1) }{\Gamma(s)}\frac{1}{(m r)^{2n}}.
\end{equation}
As expected, this is an expansion around $m \sim \mu \rightarrow \infty$, with coefficients related to the Seeley--DeWitt coefficients on $S^2$ that we have computed in Eq.~\eqref{eq:S2coeff}.
Note that the extra gamma function gives rise to a further $n!$ enhancement of the large-order divergence.
For \(s=-1/2\) we recover the grand potential:
\begin{equation}
  \label{eq:large-m-grand-potential}
    \omega(m) = - \frac{1}{2} \zeta \left( \left. - \frac{1}{2} \right| S^2 , m^2 \right) =  r^2 m^3 \sum_{n=0}^\infty \frac{ \omega_n}{(m r)^{2n}} =   \frac{1}{3} r^2 m^3 - \frac{1}{24} m  + \frac{7}{1920} \frac{1}{m r^2} + \dots 
\end{equation}
The coefficients of the grand potential can be written in closed form as
\begin{equation}
  \omega_n =- \frac{1}{4 \pi}  \sum_{k \neq 0} \frac{(-1)^k}{(k\pi)^{2n}} \Gamma\left( n + \frac{1}{2} \right) \Gamma \left( n - \frac{3}{2} \right).
  \label{eq:gran-pot-coeff}
\end{equation}
The double gamma function renders the matching with the general behavior of the higher-order trans-series coefficients in Eq.\eqref{eq:GenericLargeOrder} slightly less immediate.
We can make use of the  identity
\begin{equation}
  \begin{aligned}
	  2^{2n} \Gamma(n+\tfrac{1}{2}) \Gamma(n - \tfrac{3}{2}) &=   \sqrt{\frac{\pi}{2} }  \sum_{k=0}^{\infty} \gamma_k \Gamma\left( 2n - \frac{3}{2} - k \right)  \\
  &= \sqrt{\frac{\pi}{2} } \left[ 8 \Gamma(2n - \tfrac{3}{2}) + 15 \Gamma(2n - \tfrac{5}{2}) + \frac{105}{16}  \Gamma(2n - \tfrac{7}{2}) + \dots \right],
\end{aligned}
\end{equation}
where the coefficients $\gamma_k$ can be computed recursively.

This relation allows us to match with the generic large-order behavior Eq.~\eqref{eq:GenericLargeOrder}, where $z= 1/(mr)^2$, obtaining
\begin{align}
	\beta &= 2, & b_k&= -\frac{3}{2}, & A_k &= 2\pi k, & \frac{S_k}{2 \pi i} a_0^{(k)} &= \frac{(-1)^{k+1}}{4\sqrt{2\pi}} \frac{\gamma_0}{(2\pi \abs{k} )^\frac{3}{2}}  , & a^{(k)}_{\ell >0} &= \frac{\gamma_\ell}{\gamma_0}\frac{1}{(2\pi \abs{k})^\ell \gamma_0},
\end{align}
so that the non-perturbative corrections to the grand potential have the form
\begin{equation}
  \label{eq:non-perturbative-grand}
\omega(m )   \supset \sqrt{r m^3}   \frac{(-1)^k}{( 2 \pi \abs{k})^{\frac{3}{2}}}  e^{-(2 \pi \abs{k} ) r m}  \sum_{\ell=0}^\infty \left( \frac{\gamma_\ell}{\gamma_0 } \right) \frac{1}{(2\pi \abs{k} m r)^{\ell}}.%
\end{equation}
We find again the same structure of non-perturbative corrections that we had seen on the torus.
They are controlled by the exponential of the typical length of the manifold \(2 \pi r\) and the chemical potential \(\mu\).

The coefficients $\gamma_\ell$ are factorially growing and alternating in sign.
They can be shown to appear in Henkel's expansion of the modified Bessel function:
\begin{align}
K_2(z) &\sim \sqrt{\frac{\pi}{2z}} ea^{-z} \sum_{\ell =0}^\infty \left( \frac{\gamma_\ell}{\gamma_0} \right) \frac{1}{z^{\ell}} & \text{as \(  z \rightarrow \infty.\)}
\end{align} 
This fact will become relevant when we discuss Borel resummation.

\bigskip

\begin{figure}
  \centering
  \includegraphics[width=.7\textwidth]{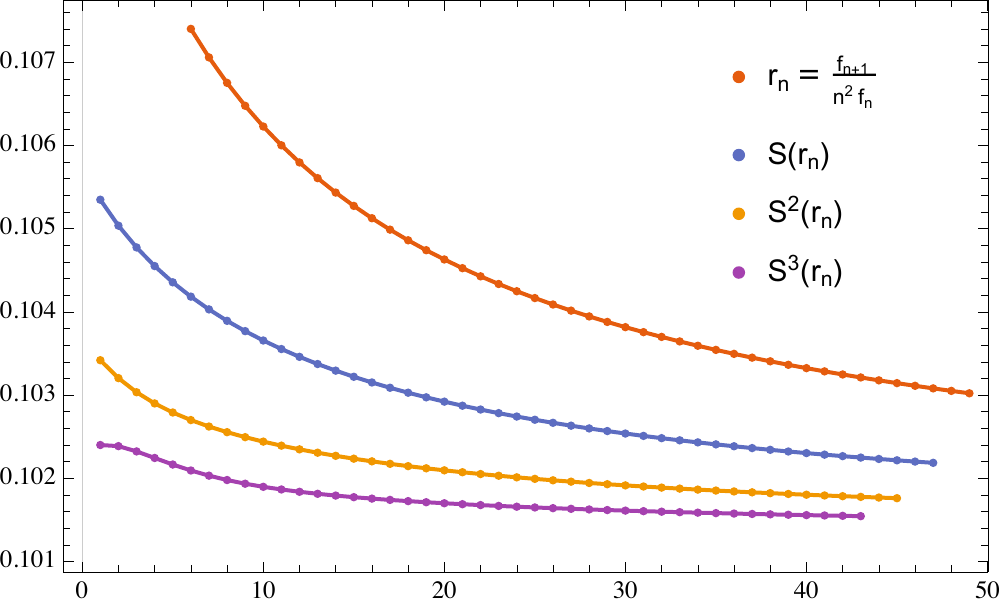}
  \caption{Ratio $r_n = n^{-2} f_{n+1} / f_n $ for the coefficients in the expansion of the free energy, together with the first three Shank transforms~\cite{shanks1955non} as function of \(n\). The convergence at large \(n\) to  a constant value of $\order{1}$ indicates a double-factorial leading behavior $f_n \sim (n!)^2$, as expected from the form of the non-perturbative contributions. }
  \label{fig:fn-coefficients}
\end{figure}

Now that we have understood the behavior of the grand potential, we can move to the free energy.
The Legendre transform can be computed order by order in \(\hat q \) starting from the perturbative part:
\begin{align}
  \label{eq:Legendre}
  \hat{q} &= \dv{\mu} \omega(\mu) & \implies &&  r m(\hat q) &=  {\hat q}^{1/2} - \frac{1}{24} {\hat q}^{-1/2} + \frac{43}{5760 }\hat q^{-3/2} + \dots\\
  f(\hat q) &= \mu \hat q - \omega(\mu) & \implies &&  f(\hat q) &= \frac{2}{3 r} \hat q^{3/2} + \frac{1}{6 r} \hat q^{1/2} - \frac{7}{720 r} \hat q^{-1/2} + \dots
                                                          \label{eq:sphere-free-energy}
\end{align}
This is clearly an asymptotic series.
For the scope of this paper, it is sufficient to consider just the leading non-perturbative terms appearing in $f(\hat{q})$ and thus in the critical exponents.
These already give a high level of precision when matching to the small-charge result, and are obtained by using the leading-order approximation of Eq.~\eqref{eq:Legendre}.
We find
\begin{equation}
	f(\hat{q})  \supset \frac{\hat{q}^{3/4}}{r}  \frac{(-1)^k}{( 2 \pi \abs{k})^{\frac{3}{2}}}  e^{-(2 \pi \abs{k} ) \sqrt{\hat{q}}} + \dots
\end{equation}
This corresponds to a $2n!$ factorial divergence of the perturbative series of $f(\hat{q})$, or equivalently of the critical exponents of the model (see Figure~\ref{fig:fn-coefficients}).

From the point of view of the \ac{eft}, the \((2n)!\) divergence is a tree-level effect.
If we identify the grand potential with an effective action (see Section~\ref{sec:lessons-from-large-N} for details), the Wilsonian coefficients form a divergent series.
This is to be compared with the \(n!\) divergence that we generically expect in \ac{qft} from the proliferation of Feynman diagrams (see Appendix~\ref{sec:lipatovs-instantons} for a discussion of Lipatov's instantons).
In the regime under discussion, the classical divergence is more important than the quantum one.

Having the explicit expressions for all the terms in the perturbative series, it is possible to extrapolate the result to arbitrarily small charge if we manage to resum the associated trans-series into a resurgent function.
This can be done using Borel resummation, as shown in the next paragraph.

\paragraph{Borel resummation.}

In the first part of this section we have constructed the general form of non-perturbative terms associated to our factorially divergent series that we have assumed to be the asymptotics of a resurgent function.
We are however still left with the problem of giving a meaning to the factorially divergent series we started with.
The Borel resummation is a prescription that achieves this goal, systematically incorporating the non-perturbative terms that we have found (see Appendix~\ref{sec:borel-transform} for a short discussion).

Let us start with the Borel transform of the heat trace expansion in Eq.~\eqref{eq:Phi0}.\footnote{We divide out the prefactor $r^2/t$ in order to work with power series with positive powers only. It can be re-introduced at the end of the analysis.}
In the case of $\mani = S^2$ we have the luxury of having a closed-form expression for the Borel transform using the definition\footnote{This definition is equivalent to the one given in Appendix~\ref{sec:borel-transform} up to the mapping $\zeta \rightarrow \zeta^2$. In this way we have a Borel transform without branch cuts.}
\begin{equation}
	\mathcal{B}\{\Phi^{(0)} \} (\zeta)=   \sum_{n=0}^\infty \frac{a_n}{\Gamma(n+1/2)} \zeta^{2n} =  \frac{1}{\sqrt{\pi}} \frac{\zeta}{\sin \zeta} ,
\end{equation}
where we have used the Taylor expansion of
\begin{equation}
  \frac{1}{\sin(z)} = 2 \sum_{n=0}^\infty  B_{2n} \frac{(-1)^n(1-2^{2n-1})}{(2n)!} z^{2n-1}.
\end{equation}
The appropriate Borel resummation in the direction $\theta = 0$ is then 
\begin{equation}
 \mathcal{S} \{ \Phi^{(0)} \} (z)=  \frac{2}{\sqrt{z}}  \int_0^\infty \dd{\zeta} \,e^{-\zeta^2/z} 	\mathcal{B}\{\Phi^{(0)} \} (\zeta) = \frac{2}{\sqrt{\pi z}} \int_0^\infty \dd \zeta\, \frac{\zeta\,e^{-\zeta^2 / z}}{\sin \zeta}.
 \label{eq:Kernel-resummation}
\end{equation}
This is the integral representation of the $S^2$ heat trace originally found in~\cite{perrin1928etude} and recovered here as a Borel integral.
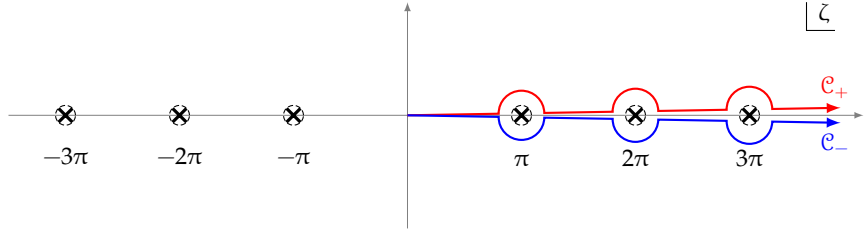
\begin{figure}
  \centering
  \begin{footnotesize}
    \begin{tikzpicture}[x=1.5cm,y=1.5cm,
      pole/.pic={
        \tikzset{scale=sin 5}
        \clip [preaction={draw, dash pattern=on 2pt off 1pt}] circle [radius=1];
        \draw [very thick] (-1,1) -- (1,-1) (-1,-1) -- (1,1);
      }]

      \draw [help lines, -latex] (-3.5,0) -- (4,0);
      \draw [help lines, -latex]  (0,-1) -- (0,1);
      \foreach \x/\y in {1/{},2/2,3/3}
      {
        \pic at (\x, 0) {pole};
        \node[label={\(\y \pi\)}] at (\x,-.6){};
        \pic at (-\x, 0) {pole};
        \node[label={\(-\y \pi\)}] at (-\x,-.6){};
      }

      \draw[thick, red,-latex, rotate=1] (0,0) -- (8/10, 0) arc (180:0:1/5) -- (12/10, 0) -- (18/10, 0)  arc (180:0:1/5) -- (22/10, 0) -- (28/10, 0)  arc (180:0:1/5) -- (32/10, 0) -- (38/10, 0);
      \node[red] at (3.75,.25) {\(\mathcal{C}_+\)};

      \draw[thick, blue,-latex, rotate=-1] (0,0) -- (8/10, 0) arc (180:360:2/10) -- (12/10, 0) -- (18/10, 0)  arc (180:360:2/10) -- (22/10, 0) -- (28/10, 0)  arc (180:360:2/10) -- (32/10, 0) -- (38/10, 0);
      \node[blue] at (3.75,-.25) {\(\mathcal{C}_-\)};
      
      \draw (3.5,1) -- (3.5,.75) -- (3.75,.75);
      \node at (3.65,.9) {\(\zeta\)};
    \end{tikzpicture}

  \end{footnotesize}
  \caption{Pole structure and integration contours \(\mathcal{C}_\pm\) for the lateral Borel transform of the trace of the heat kernel on \(S^2\). The two choices differ by the residues at \(y = k \pi \), \(k = 1, 2, \dots \).}
  \label{fig:lateral-Borel-sphere}
\end{figure}

The Borel integral however is ill-defined: the integrand has simple poles on the integration path for \(\zeta = k \pi\), \(k = \in \setZ^+\), which indicate that the series is not Borel summable and $\theta = 0$ is a Stokes line.
We thus define lateral Borel transforms $\mathcal{S}_\pm \{ \Phi^{(0)} \} (t)$, whose integration contours pass over or under the poles  (see Figure~\ref{fig:lateral-Borel-sphere}).
This introduces the following ambiguity in the summation:
\begin{equation}
\begin{aligned}
	 [\mathcal{S}_+ - \mathcal{S}_{-} ] \{ \Phi^{(0)} \} (z) &= - (2 \pi i)  \sum_{k=1}^\infty \Res_{\zeta = k \pi} \left( \frac{2}{\sqrt{\pi z}}  \frac{\zeta\,e^{-\zeta^2 /z}}{\sin \zeta} \right) \\ 
    &= 2 i z \pqty{\frac{\pi}{z} }^{3/2}  \sum_{k\neq0}^\infty (-)^{k+1}  \abs{k} \,  e^{- k^2 \pi^2/z } ,
\end{aligned}
\end{equation}
which matches exactly with Eq.~(\ref{eq:non-perturbative-heat}).
Each term corresponds to a pole on the positive real axis in the Borel plane.
For either choice of the contour, the path moves away from the real axis and the integral  picks up an imaginary contribution. Thus, the (ambiguous) trans-series associated to the heat trace is
\begin{equation}
 \Tr \left[ e^{\left(\Laplacian - \tfrac{1}{4 r^2}\right) t} \right]  =  \frac{2 r^3}{\sqrt{\pi} t^{3/2}} \int_{\mathcal{C}_\pm} \dd{\zeta}  \frac{\zeta\, e^{-\zeta^2 r^2/t}}{\sin \zeta} + 2i \left( \frac{\pi r^2}{t} \right)^{\frac{3}{2}} \sum_{k\neq 0} \sigma_k^{\pm} (-1)^{k+1}  \abs{k}  e^{- \frac{k^2 \pi^2 r^2}{t}}.
\label{eq:heat-trans-series}
\end{equation}
An analogous treatment can be carried out for the grand potential starting from the perturbative coefficients in Eq.\eqref{eq:gran-pot-coeff}.
However, the Borel transform of $\omega^{(0)}$ does not have a neat closed form.
Instead, we apply the Mellin transform to the Borel resummation of the heat kernel in Eq.~\eqref{eq:Kernel-resummation} to find
\begin{equation}
\begin{aligned}
	\zeta_\pm(s | S^2 , m^2 ) 
	&=  \frac{r^{2s}}{ \Gamma(s)} \int_0^\infty \dd{z} z^{s-2} e^{-m^2r^2 z} \left[  \mathcal{S}_{\pm} \{ \Phi^{(0)}\} (z) - 1 - \frac{z}{12} \right] +   \frac{ r^2 m^{2-2s} }{s-1} + \frac{m^{-2s}}{12} .
	\end{aligned}
\end{equation}
The Mellin integral has been analytically continued so that it is convergent for $s = -1/2$. This allows us to exchange the integration order and obtain
\begin{equation}
  \label{eq:resurged-grand-potential}
	\begin{aligned}
	\omega_{\pm}(m) &= - \frac{1}{2} \zeta \left(\left. - \frac{1}{2} \right| S^2 , m^2 \right) \\
	&=   \frac{1}{3} r^2 m^{3} - \frac{m}{24}  + \frac{m^2 r}{\pi}  \int_{\mathcal{C}^\pm} \frac{ \dd \zeta}{\zeta^2} \left( \frac{\zeta}{\sin \zeta} - 1 - \frac{\zeta^2}{6} \right) K_2(2 m r \zeta).\\
    \end{aligned}
\end{equation}
From this expression we can identify the (non-standard) Borel resummation of the series $\omega^{(0)}$ with the integral 
\begin{equation}
	\mathcal{S}_{\pm}\{\omega^{(0)}\} =  \int_{\mathcal{C}^\pm} \frac{ \dd \zeta}{\zeta^2} \left( \frac{\zeta}{\sin \zeta} - 1 - \frac{\zeta^2}{6} \right) K_2(2 m r \zeta).
\end{equation}
It has a discontinuity, 
\begin{equation}
  \label{eq:grand-potential-residues}
[	\mathcal{S}_{+}-\mathcal{S}_-]\{\omega^{(0)}\} = \sum_{k=1}^\infty \frac{(-1)^k}{k^2 \pi^2} K_2(2 \pi k mr),
\end{equation}
which turns out to be in perfect agreement with the result that we had found above in Eq.~\eqref{eq:non-perturbative-grand}. 
 In the analysis of the heat kernel we had found that the non-perturbative contributions are semiclassically exact, \emph{i.e.} consist of only one term.
The same applies also to the grand potential: we still have only one term but the non-perturbative corrections contain a Bessel function instead of the usual instanton-like exponentials typical of \ac{qft} problems.

\bigskip

The non-perturbative ambiguities from the Borel summation in the grand potential and the free energy are related to the ones appearing in the heat trace in Eq.~\eqref{eq:heat-trans-series}.
They can be fixed in different ways:
\begin{itemize}
\item Imposing the reality of the heat trace in Eq.~\eqref{eq:heat-trans-series}  for $ t \in \setR^+$.
  In general this is not guaranteed to fix completely the parameters $\sigma_k$, but in our case it turns out to be sufficient (see Appendix~\ref{sec:Dawson}) and one finds $\sigma_k^{\pm} = \pm 1/2$, which implies $S_k = 1 $ for all values of \(k\).
  The heat trace then reads
	\begin{equation}
	\begin{aligned}
	 \Tr \left[ e^{\left(\Laplacian - \tfrac{1}{4r^2}\right) t} \right]  &=  \frac{2}{\sqrt{\pi}} \left( \frac{r^2}{t} \right)^{\frac{3}{2}} \int_{\mathcal{C}_\pm} \dd{\zeta}  \frac{\zeta\, e^{-\zeta^2 r^2/t}}{\sin \zeta}  \pm i \left( \frac{\pi r^2}{t} \right)^{\frac{3}{2}} \sum_{k\neq 0} (-1)^{k+1}  \abs{k}  e^{- \frac{k^2 \pi^2 r^2}{t}} \\
	 &=  \frac{2}{\sqrt{\pi}} \left( \frac{r^2}{t} \right)^{\frac{3}{2}} \operatorname{P.V.}\left[  \int_{\mathcal{C}_\pm} \dd{\zeta}  \frac{\zeta\, e^{-\zeta^2 r^2/t}}{\sin \zeta} \right] ,
	 \label{eq:final-heat-trace}
	 \end{aligned}
	\end{equation}
    which is unambiguous and real, despite the appearances. This usually holds for various systems involving  \acp{ode}~\cite{Aniceto:2013fka}. Even though the heat trace is the solution of a non-linear \ac{pde} (the heat equation) at coincident points, it is interesting to see how its trans-series structure can be deduced from a linear Dawson's \ac{ode}.
    \item Finding a (path-)integral definition of the heat trace, where a trans-series structure arises automatically from the semiclassical expansion around non-trivial saddle points. For ordinary integrals it has been shown in~\cite{Cherman:2014ofa} that such a reality prescription is sufficient as it yields the Lefschetz thimble decomposition of the integral. The same authors have shown that in the case of path integrals, also unstable saddles play a role in the cancellation of ambiguities. We will pursue this direction further in the next section and show that a similar phenomenon arises in a path integral formulation of the heat trace.
\end{itemize}

\section{Worldline interpretation}%
\label{sec:worldline}

It has been known since the work of Feynman that Green's functions of elliptic operators can admit a representation in terms of quantum mechanical path integrals.
In the worldline formalism, see~\cite{Schubert:2001he} and references therein, one constructs an appropriate quantum mechanics path integral which computes the determinant of a given operator.
It has been successfully applied to the computation of \ac{qft} amplitudes and effective actions on classical field backgrounds. 
Heat kernels can be represented as the worldline integral of a free particle moving on the curved manifold of interest. However, quantum mechanics on curved space is not a trivial subject, and has been known since DeWitt~\cite{DeWitt:1957at} to be plagued with ambiguities related to the problem of defining path-integral measures on curved space.
Most of the difficulties have been solved in the past years and have lead to a perturbative definition of such path integrals, see for example~\cite{DeBoer:1995hv}, which have been shown to match the first few Seeley--DeWitt coefficients on general manifolds~\cite{Bastianelli:2006rx}.

In this section we will use the worldline approach to show that the trans-series Eq.~\eqref{eq:final-heat-trace} can be obtained as a saddle-point approximation as $t \rightarrow 0^+$ of an appropriate quantum mechanical path integral computing the heat trace. 
This results in an entirely geometric interpretation of the non-perturbative terms and ambiguities that appear in the resurgent analysis.
This same structure carries over from the heat trace to the grand potential $\omega$, and ultimately to the large-charge expansion of the conformal dimension in the double-scaling limit of the $O(2N)$ model.

\subsection{The heat kernel as a path integral}
\label{sec:path-integral}

The starting point of the worldline approach to the calculation of functional determinants is Schwinger's representation:
\begin{equation}
	 \log(\det(- \del_0^2{} - \Laplacian{} + \mu^2 )) = -  \int_0^\infty \frac{\dd{t}}{t} e^{-\mu^2 t} \Tr(e^{(\del_0^2{} + \Laplacian{}) t} ) .
\end{equation}
This functional determinant computes the grand potential \(\omega(\mu)\), see Appendix~\ref{sec:largeN}.

On a product manifold such as $\setR \times \mani$ the trace in the Schwinger integral factorizes, so that one can study directly the heat trace on $\mani$:
\begin{equation}
   \log(\det(- \del_0^2{} - \Laplacian{} + \mu^2 )) =  -  \int_0^\infty \frac{\dd{t}}{t} e^{-\mu^2 t} \frac{1}{\sqrt{4 \pi t}}  \Tr(e^{\Laplacian{} t} ).
\end{equation}
The idea is to interpret the heat trace as the partition function for a particle at inverse temperature \(t\) and Hamiltonian \(H = - \Laplacian{}\), \emph{i.e.} a free quantum particle moving on $\mani$ ~\cite{DeWitt:2003pm,Camporesi:1990wm,Bastianelli:2006rx}.
If we take a coordinate system $x^\mu$ on $\mani$ then the classical action of the free particle is 
\begin{equation}
  S[X] = \frac{1}{4} \int_0^t \dd{\tau} g_{\mu \nu}(x) \dot x^\mu(\tau) \dot x^\nu(\tau),
\end{equation}
where \(g_{\mu\nu}\) is the metric on $\mani$ and $x^\mu:(0,t) \to  \mani$ is the worldline described by the motion of the free particle. 
The heat trace is then related to a path integral over closed loops (Feynman--Kac formula) of the form%
\footnote{The action has a reparametrization invariance which we have fixed, which appears as a gauge-invariance of the heat trace. It will not play a role in our computation.} 
\begin{equation}
  \Tr \left[ e^{ \Laplacian{} t} \right] \equiv \int\displaylimits_{x(t) = x(0)} \DD{x^\mu} e^{-S[x]}.
  \label{eq:working-definiton}
\end{equation}
We will take this path integral as the definition of the heat trace.
Note however that the application of the Feynman--Kac formula is quite involved because of the intrinsic diffeomorphism invariance and of the ordering ambiguities introduced by the curvature terms in the quantization of the Hamiltonian. 
All these issues have been resolved for the semiclassical expansion of Eq.~\eqref{eq:working-definiton} around the trivial loop $x_{\rm cl}^\mu(\tau) = 0$.

In this work we take advantage of the fact that all these modifications are subleading in the expansion as $t \rightarrow 0^+$.
This  is in fact a semiclassical expansion: after rescaling the worldline time $\tau \rightarrow \tau t $, the action can be rewritten as
\begin{equation}
  S[x] = \frac{1}{4t} \int_0^1 \dd{\tau} g_{\mu \nu}(x) \dot x^\mu \dot x^\nu ,
\end{equation}
and the small-$t$ expansion of the heat kernel corresponds to the expansion in $\hbar$ in this quantum mechanical system.
In this limit, the path integral localizes around the saddle points of \(S[x]\) and we can expand it perturbatively in powers of \(t\).

The Euler--Lagrange equations for our action are the geodesic equations
\begin{equation}
  \ddot x_{\text{cl}}^\mu + \Gamma^\mu_{\nu \rho}(x) \dot x_{\text{cl}}^\nu \dot x_{\text{cl}}^\rho = 0,
\end{equation}
so that the heat trace path integral localizes on a sum over all the closed geodesics \(\gamma\) on \(\mani\).
These non-trivial geodesics are the equivalent of the worldline instantons in~\cite{Dunne:2006st,Dunne:2005sx} which govern the non-perturbative contributions to Euler--Heisenberg-type Lagrangians~\cite{Dunne:2004nc}.

Ordinary instanton calculus shows that, in general, each of these saddles will come with its own perturbative series in \(t\), weighted by \(e^{-\ell(\gamma)^2/(4t) }\), where \(\ell(\gamma)\) is the length of \(\gamma\), so that the semi-classical expansion has the general form
\begin{multline}
  \Tr \left[ e^{  \Laplacian{} t} \right]  = t^{-b_0} \sum_{n=0}^\infty a_n^{(0)} t^{n} + \sideset{}{'} \sum_{\text{\(\gamma \in \) closed geodesics}} e^{-\frac{\ell(\gamma)^2}{4 t} } t^{-b_\gamma} \sum_{n=0}^\infty a_n^{(\gamma)} t^{n},
  \label{eq:semiclassical}
\end{multline}
where the sum runs over the non-trivial geodesics, and the \(b_\gamma\) depend on the geometry.
The sequences $a_n^{(\gamma)}$ are generally expected to be factorially growing by usual Feynman graph proliferation arguments. 

The similarity with the structure of the generic trans-series in Eq.~\eqref{eq:trans-series} is not a coincidence: the latter were introduced to match semiclassical expansions, where they appear naturally. 
However, there is a  conceptual difference with respect to the resurgent analysis carried on in the previous section.
Resurgence does not rely on the existence of a non-perturbative definition of the observable we want to compute (such as a path integral definition).
For this reason there is in general no geometric interpretation of the trans-series structure and there are ambiguities which cannot be fixed a priori.
In our case we have shown that the ambiguities can be removed by just imposing the reality of the end result.
In the following we will see that the path integral definition reproduces the result Eq.~\eqref{eq:final-heat-trace} without further input.
\subsection{The torus}
\label{sec:torus-path-integral}

In this section we review the derivation of the heat trace on $T^2$ that we have found in Eq.~\eqref{eq:Torus-trace-Poisson} via a path integral prescription. 

Consider a square torus of side \(L\), with metric%
\begin{equation}
  \dd s^2 = g_{ij} \dd{x^i}  \dd{x^j} = (\dd{x^1})^2 + (\dd{x^2})^2.
\end{equation}
Since the torus is flat, no subtleties related to curved space arise and our working definition in Eq.~\eqref{eq:working-definiton} is correct.
The heat trace is defined via the path integral
\begin{equation}
  \Tr(e^{\Laplacian{} t})   =  \int\displaylimits_{x(t)=x(0)} \DD{x} e^{- \frac{1}{4t} \int_0^1 \dd{\tau} ( (\dot x^1)^2 + (\dot x^2)^2) }.
\end{equation}
In the limit \(t \to 0^+\) it can be computed semiclassically via saddle-point approximation.

\begin{figure}
  \hfill
  \begin{tikzpicture}[scale=.8]
    \coordinate (Origin)   at (0,0);
    \coordinate (XAxisMin) at (-1,0);
    \coordinate (XAxisMax) at (6.5,0);
    \coordinate (YAxisMin) at (0,-1);
    \coordinate (YAxisMax) at (0,3.5);
    \draw [thin,-latex] (XAxisMin) -- (XAxisMax);%
    \draw [thin,-latex] (YAxisMin) -- (YAxisMax);%

    \clip (-.6,-.6) rectangle (6.6,3.6); %

    \draw[thin,gray] (-2,-2) grid[step=1] (7,7);

    \coordinate (Xone) at (1,0);
    \coordinate (Xtwo) at (0,1);

    \filldraw[fill=gray, fill opacity=0.3, draw=black] (Origin)
    rectangle ($(Xone)+(Xtwo)$);

    \draw[thick, red, -latex] (0,0) -- (5,2);
    \draw[thick, blue, -latex] (0,0) -- (1,3);
  \end{tikzpicture} \hfill
  \includegraphics[width=.5\textwidth]{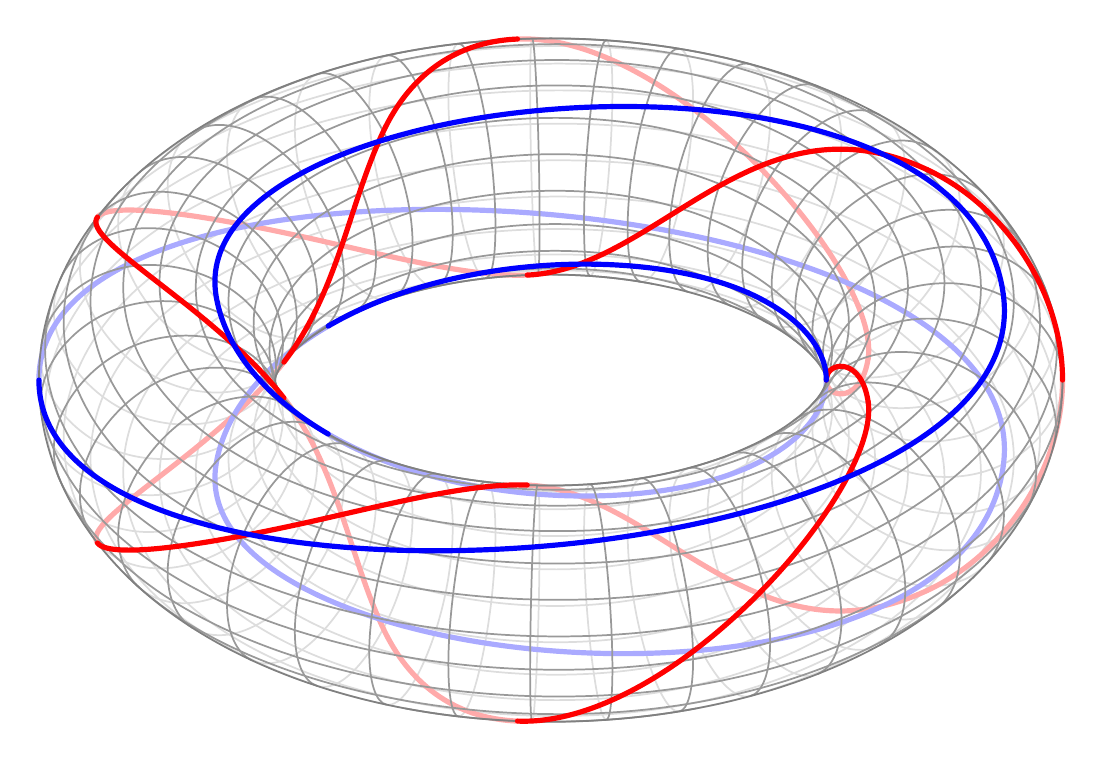}
  \hfill
  \caption{Closed geodesics on the torus labeled by the integers \((1,3)\) (blue) and \((5,2)\) (red) as segments in \(\setR^2\) and on a three-dimensional rendering.}
  \label{fig:torus-geodesics}
\end{figure}
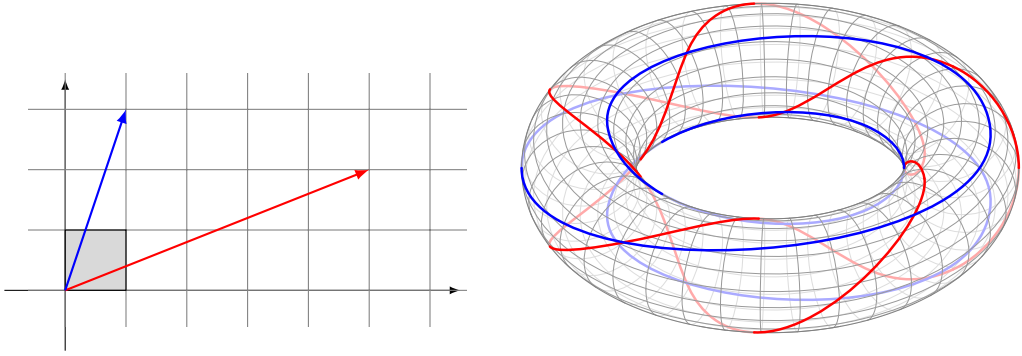

We represent the torus as  \(\setR^2\) with the identifications \(x^i \simeq x^i + L\).
If we fix a point, say the origin, we obtain a lattice \(\setZ^2 \) of equivalent points.
The closed geodesics passing through this point are straight lines joining the origin with another point of the lattice (see Figure~\ref{fig:torus-geodesics}).
This means that closed geodesics can be labeled by pairs of integers \((k_1, k_2)\) (this includes the trivial geodesic of zero length).
The corresponding length is simply
\begin{equation}
  \ell(k_1, k_2) = L \sqrt{k_1^2 + k_2^2}.
\end{equation}

The field in the path integral can be decomposed as the sum of the classical solutions \(X^i_{\text{cl}}\) and fluctuations \(h^i(\tau)\),
\begin{equation}
  X^i(\tau ) = X^i_{\text{cl}}(\tau) + h^i(\tau) = k^i L \tau + h^i(\tau) .
\end{equation}
The action is just quadratic since $T^2$ is flat, and we can separate the two contributions:
\begin{equation}
  S[X] = S[X_{\text{cl}}] + S[h] = \frac{L^2 (k_1^2 + k_2^2)}{4 t} + \frac{1}{4t} \int_0^1 \dd{\tau} \left[ (\dot h^1)^2 + (\dot h^2)^2 \right] .
\end{equation}
Then the path integral becomes
\begin{equation}
  \Tr(e^{\Laplacian{} t})   = \int_{T^2} \dd{x}  \int\displaylimits_{X(1)=X(0)=x} \DD{X} e^{S[X] } =L^2 \sum_{\mathbf{k} \in \setZ^2} e^{-\frac{L^2 (k_1^2 + k_2^2)}{4 t}} \int\displaylimits_{h(t) = h(0) = 0} \DD{h} e^{-S[h]},
\end{equation}
where we have added the integral over the fixed point \(x\) through which each geodesic passes.
The remaining path integral is Gaussian and can be defined, for example in zeta-function regularization, up to a normalization constant which is chosen to reproduce the standard worldline normalization
\begin{equation}
 \int\displaylimits_{h(t) = h(0) = 0} \DD{h} e^{- \frac{1}{4t} \int_0^1 \dd \tau \dot{h}^2} = \frac{1}{\sqrt{4\pi t}}.
 \label{eq:normalisation-factor}
\end{equation}
The final result is
\begin{equation}
 \Tr(e^{\Laplacian{} t})  = \frac{L^2}{4\pi t} \sum_{\mathbf{k} \in \setZ^2} e^{- \frac{L^2 \norm{\mathbf{k}}^2}{4t}} ,
\end{equation}
which is precisely the right-hand side of the Poisson resummation formula used in Eq.~(\ref{eq:Torus-trace-Poisson}).

This correspondence is known in the literature as spectrum-geodesic duality on compact manifolds.
For each eigenvalue of the Laplacian there is a corresponding closed geodesic, see~\cite{Camporesi:1990wm,Grosche:1995ria} for a review.
There exist (Selberg-like) trace formulas which realize this duality for the heat trace, and the right hand-side can be interpreted as a saddle-point approximation of an appropriate path integral.
In the torus case, the duality is realized via Poisson resummation.

\subsection{The sphere}
\label{sec:sphere-path-integral}

In the case of $\mani = S^2$ the subtleties related to curved space play a role. As mentioned at the beginning of the section, we will employ a ``natural'' generalization of flat space path integrals and define
\begin{equation}
  \mel{y}{e^{ \Laplacian{} t }}{x} \equiv \int_{x(0)=x}^{x(t)=y} \DD{x^\mu} e^{- \frac{1}{4t } \int_0^1 \dd \tau \, g_{\mu\nu}(x) \dot{x}^\mu \dot{x}^\nu}.
\label{eq:path-integral-def}
\end{equation}
We will use standard polar coordinates on the sphere $x^\mu = (\theta,\phi)$ so that the volume element is $\DD{x^\mu} = \sin(\theta)\DD{\theta} \DD{\phi}$.
These will be our worldline fields, with action and \ac{eom} given by:%
\begin{align}
  S &= \frac{r^2}{4t} \int_0^1 \dd{\tau} \left[ \dot{\theta}^2 + \sin^2 \theta \dot \phi^2 \right], \\
  &
    \begin{cases}
      \ddot{\phi} + 2 \cot(\theta) \dot{\theta} \dot{\phi} = 0, \\
      \ddot{\theta} - \dot{\phi}^2 \sin(2 \theta) = 0. 
    \end{cases}
	\label{eq:action-EOM}
\end{align}	
As we are ultimately interested in the heat trace we consider coincident endpoints $x^\mu(0) = x^\mu(1) = ( \pi/2 , 0)$.\footnote{Note that the action Eq.~\eqref{eq:path-integral-def} in these coordinates is \emph{not} rotationally invariant in $\theta$ as there are two singular points (the poles). However, this appears only at higher order in $t$, so that the leading order in $t \rightarrow 0^+$ is fine. }

There are infinitely many winding geodesics which solve the \ac{eom} which are parametrized as
\begin{align}
	\theta_{\rm cl}(\tau) &= \pi/2 , &	\phi_{\rm cl}(\tau) &= 2 \pi k \tau , & k &\in \setZ,
\end{align}	
where we allow $\phi$ to extend beyond its domain with the identification $\phi \sim \phi + 2\pi$.
If we  introduce the fluctuations $h_\theta, h_\phi$ around these solutions,  satisfying Dirichlet boundary conditions, we can formally rewrite the heat kernel at coincident points as
\begin{equation}
 \mel{x}{e^{\left( \Laplacian{}  \right)t }}{x} = e^{- \frac{(2\pi k r)^2}{4 t} } \int\displaylimits_{h_i(t) = h_i(0) = 0}  \DD{h_\theta} \DD{h_\phi}  e^{  - \frac{r^2}{4t} \int_0^1 \dd \tau \left[ \dot{h}_\theta^2 - (2\pi k)^2 h_\theta^2 + \dot{h}_\phi^2 + \order{h^3} \right]   }.
\end{equation}

Standard loop-diagram counting arguments show that the $\order{h^3}$ (the ``interaction terms'') gives contributions that are higher order in $t$.
The leading order is obtained by computing the appropriate functional determinant from the quadratic action of the fluctuations.  

The integral in $h_\phi$ is massless and reproduces our normalization convention~\eqref{eq:normalisation-factor} with the substitution $t \rightarrow t /r^2$.

The integral over $h_\theta$, on the other hand, contains a zero-mode and multiple negative modes.
To see that, we expand the fluctuations in an orthonormal basis of modes satisfying
\begin{align}
	h_\theta(\tau) &= \sum_{n=1} c_{n} h_\theta^n (\tau) ,& - \frac{r^2}{2t} \left[ \partial_\tau^2 +(2\pi k)^2 \right] h_\theta^n(\tau) &= \lambda_n  h_\theta^n(\tau),
\end{align}
which can be written explicitly as
\begin{align}
	h_\theta^n &= \sqrt{2} \sin ( \pi n \tau), & \lambda_n &= \frac{\pi^2r^2}{2t} \pqty{n^2 - 4 k^2}.
\end{align}
The integration measure is then written in terms of Fourier modes as
\begin{equation}
	\int \DD{h_\theta} \equiv \prod_{n=1}^\infty \int \frac{\dd{c_n}}{\sqrt{2\pi}}.
\end{equation}

\begin{figure}
  \hfill
  \includegraphics[width=.3\textwidth]{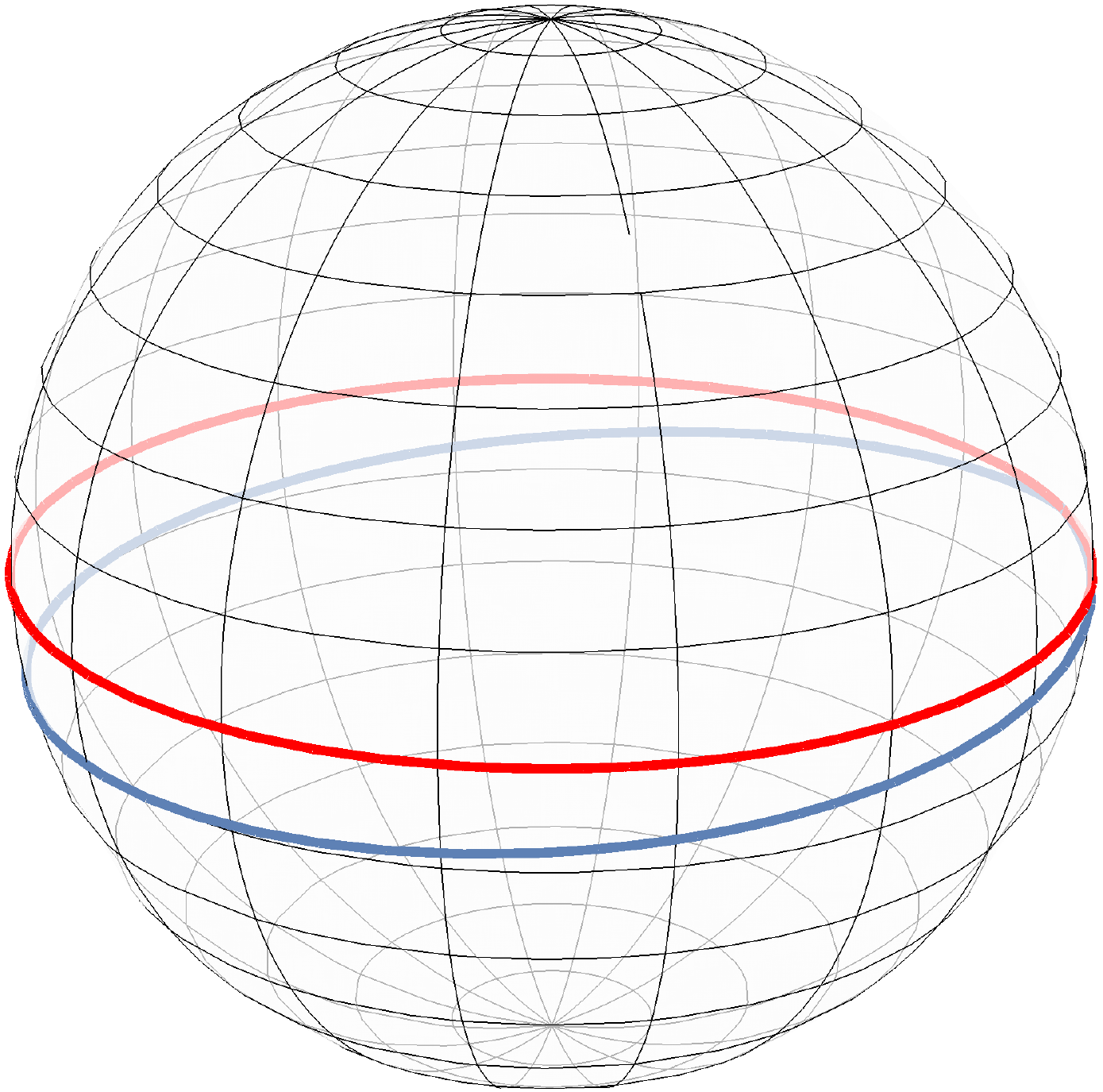}
  \hfill
    \includegraphics[width=.3\textwidth]{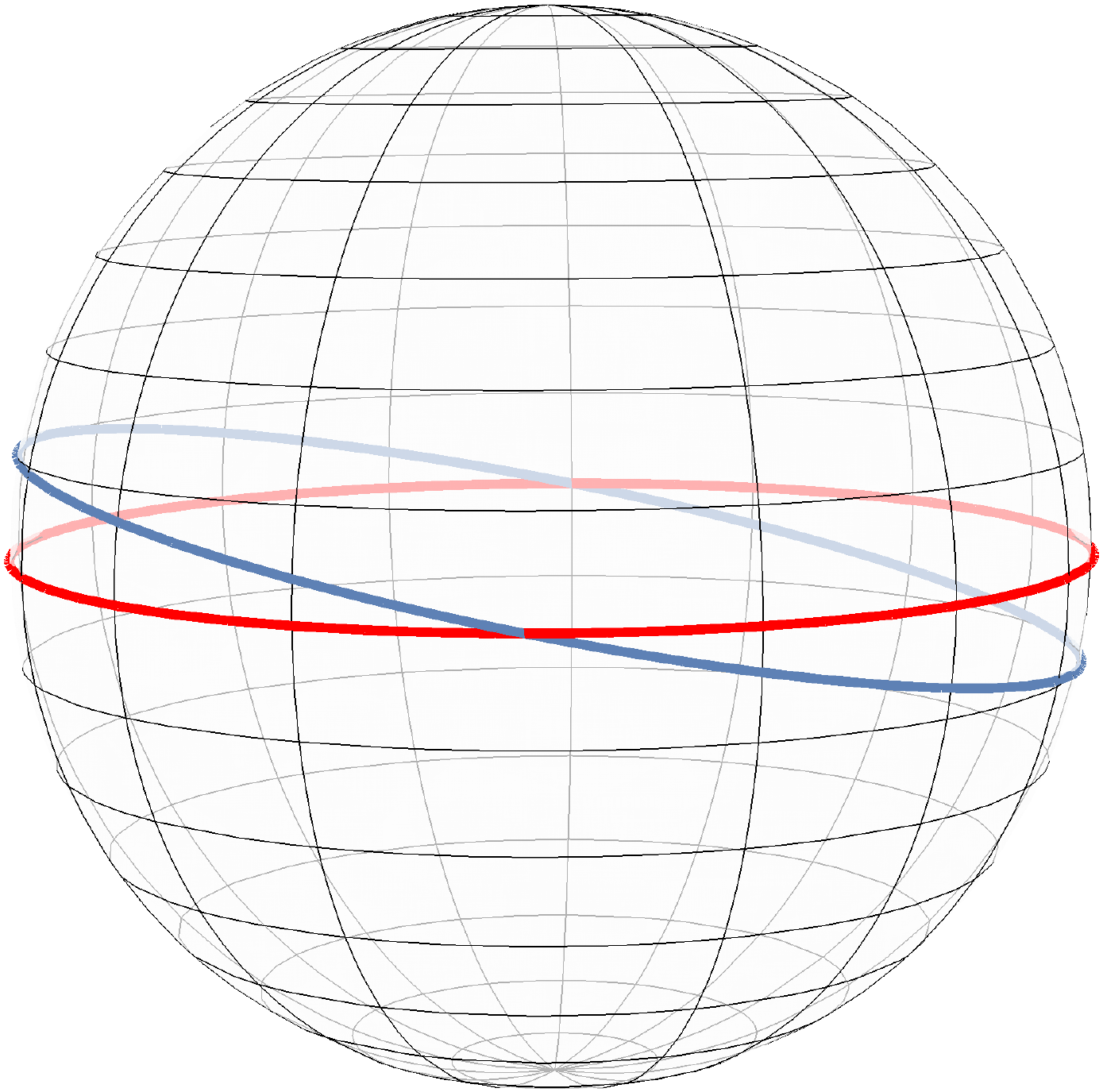}
  \hfill
  \includegraphics[width=.3\textwidth]{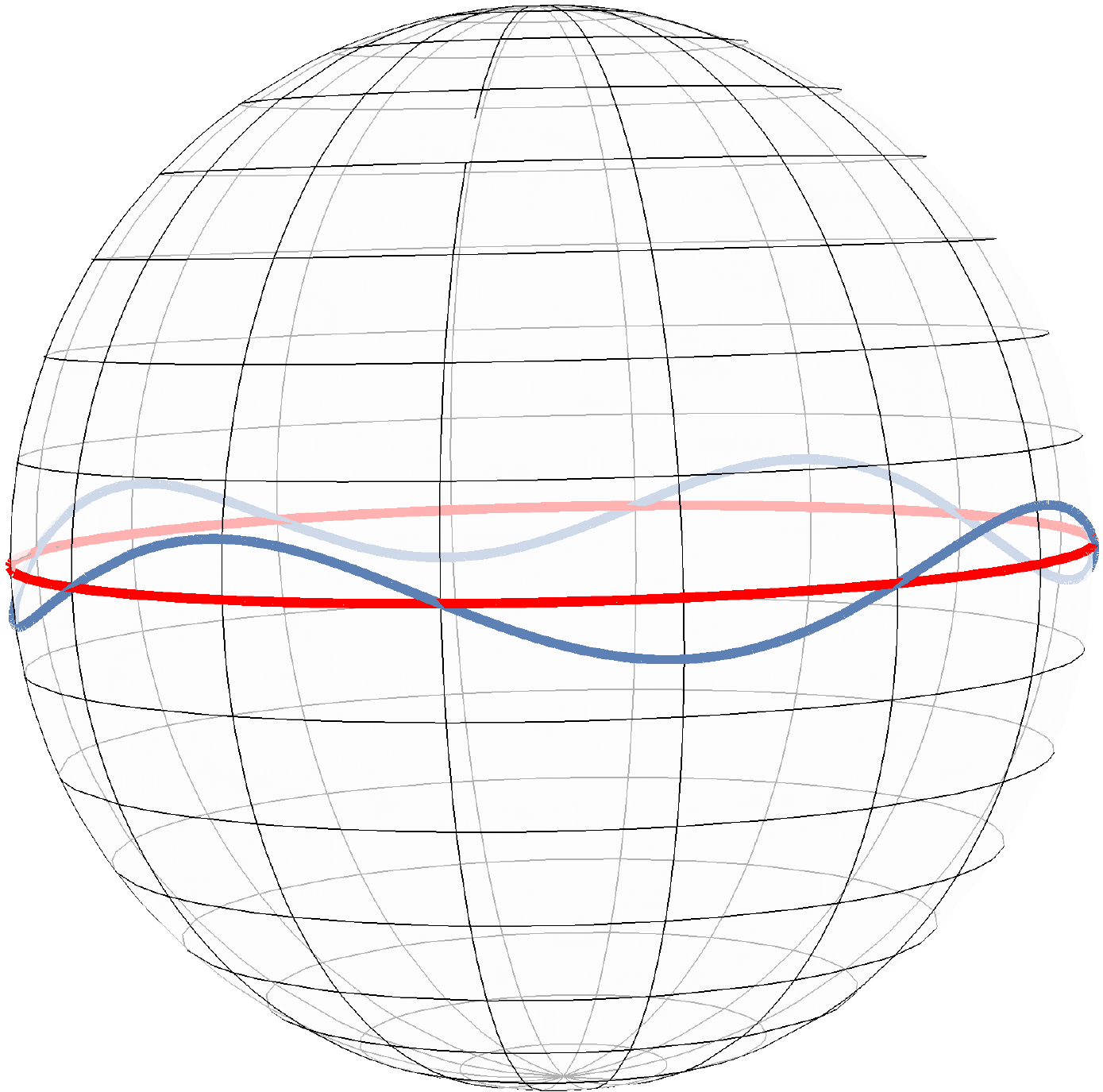}
  \caption{Unstable mode \(n=1\) (left), zero mode \(n=2\) (middle) and massive mode \(n=8\) (right) for the fluctuations around a geodesic winding once around the sphere (red).}
  \label{fig:unstable-mode-sphere}
\end{figure}

For \(n = 2k\) there is  a zero mode that needs to be treated separately, while the $(2k-1)$ modes $h_\theta^{n<2k}$ are tachyonic. 
The presence of negative modes is not surprising, since $\pi_1(S^2) = 0$. These winding geodesics are not stable as they can be contracted to a point (see Figure~\ref{fig:unstable-mode-sphere}), in contrast to what happened in the torus case, where winding geodesics are topologically stable saddles.

The zero mode corresponds to a rigid rotation of the sphere, which is a symmetry of the quadratic action.
In fact, it is easy to see that 
\begin{equation}
  \begin{cases}
    \theta^\alpha_{\rm cl} = \frac{\pi}{2} + \alpha \sin ( 2 \pi k \tau), \\
    \phi_{\rm cl} = 2\pi k \tau,
  \end{cases}
\end{equation}
is a family of solutions of the \ac{eom}~\eqref{eq:action-EOM} at leading order in the modulus $\alpha \in (0,\pi)$.
This is exactly the fluctuation $h_\theta^{2k}$, and using the rules of instanton calculus we can trade the integral over the mode amplitude $c_{2k}$ for an integral over the modulus $\alpha$, which parametrizes  the family of solutions:
\begin{equation}
	\int \frac{\dd{c_{2k}}}{\sqrt{2\pi}} = \sqrt{\frac{1}{2}} \int \frac{\dd{\alpha}}{\sqrt{2\pi}} =  \frac{\sqrt{ \pi }}{2}.
\end{equation}
The integral over the remaining modes produces the following functional determinant:
\begin{equation}
\sideset{}{'}\det \left( - \frac{r^2}{2t} \left( \partial_\tau^2 + (2\pi k)^2 \right)  \right)^{-\frac{1}{2}} = \frac{\sqrt{2} \pi \abs{k} r}{ \sqrt{t}} \det \left( - \frac{r^2}{2 t} \partial_\tau^2 \right)^{-\frac{1}{2}} \sideset{}{'}\det \left(  \text{Id} + \frac{4\pi^2 k^2}{ \partial_\tau^2}    \right)^{-\frac{1}{2}} ,
\end{equation}
where we divided and multiplied by the $n = 2k$ eigenvalue of $\partial_\tau^2$ in order to recover our initial normalization.%
\footnote{No multiplicative anomaly arises in this determinant splitting~\cite{Monin:2016bwf}.}
The remaining determinant does not need regularization and reads
\begin{equation}
	\sideset{}{'}\det \left(  \text{Id} + \frac{4\pi^2 k^2}{\partial_\tau^2}    \right)^{-\frac{1}{2}}  = e^{\frac{i\pi \nu_{q}}{2}} \left( \prod_{\substack{n =0\\n \neq 2k}}^{\infty} \abs{ 1 - \frac{4k^2}{n^2}}\right)^{-\frac{1}{2}} = \sqrt{2} \,e^{\frac{i\pi \nu_{q}}{2} },
\end{equation}
where the infinite product can be evaluated for \(k>0\):
\begin{equation}
	\prod_{n \neq 2k } \abs{ 1 - \frac{4k^2}{n^2}} = \prod_{n=1}^{2k-1} \left( \frac{4k^2}{n^2} - 1 \right) \prod_{n=2k+1}^{\infty} \left( 1 - \frac{4k^2}{n^2} \right) = \frac{ \Gamma(4k)}{2k \Gamma(2k)^2 } \frac{\Gamma(2k+1)^2}{\Gamma(4k+1)} = \frac{1}{2}.
	\label{eq:func-det}
\end{equation}
We have introduced the Morse index $\nu_q$ which arises naturally for functional determinants with $q$ negative modes~\cite{Horvathy:2007nu}, and can be interpreted as an analog of the intersection numbers arising in the Lefschetz thimble decomposition of ordinary integrals. In our computation, it is related to the fact that Gaussian integrals for negative modes have a two-fold ambiguity in their analytic continuation:
\begin{equation}
	\int \frac{\dd c_n}{\sqrt{2\pi}}\, e^{ \frac{1}{2} \lambda_n c_n^2 } = e^{\pm i \frac{\pi}{2}}  \frac{1}{\sqrt{\lambda_n}}.
	\label{eq:morse}
\end{equation}
In the computation of the determinant of Eq.~\eqref{eq:func-det} we have factored out these $q = 2k-1$ independent phases in order for each term in the infinite product to be positive definite.
The choice of each of these phases is independent for each negative mode.
We choose the same continuation procedure for each of them, so that
\begin{equation}
e^{\frac{i \pi \nu_q}{2}} = (\pm i)^{2k-1} = \mp i (-1)^k.
\end{equation}

Putting all these results together, we obtain the saddle-point approximation of the heat trace:
\begin{equation}\label{eq:pathIntHeatTrace}
	\Tr[ e^{\Laplacian{} t} ] = \frac{r^2}{t} \pqty{1 + \order{t}} \pm i \left( \frac{\pi r^2}{t} \right)^{\frac{3}{2}} \sideset{}{'}\sum_{k \in \setZ} (-1)^{k+1} \abs{k} e^{- \frac{k^2\pi^2 r^2}{t} } \pqty{1 + \order{t}}.
\end{equation}
This expression matches exactly the unambiguous trans-series representation in Eq.~\eqref{eq:final-heat-trace}.

Our prescription does not allow us to compute next-to-leading order contributions within each sector, which would also include the conformal coupling.
However, there is a prescription that computes the Seeley--DeWitt coefficients in the perturbative sector via a diagram expansion, see for example~\cite{Bastianelli:2017wsy}.
To our knowledge, there has been no attempt to reproduce the fluctuations around non-trivial sectors.
While no fundamental obstructions are expected, to obtain a one-loop exact structure of the form suggested by the non-perturbative corrections in Eq.~\eqref{eq:final-heat-trace} would require highly non-trivial cancellations in the absence of supersymmetry. 

While this is beyond the scope of this work, the matching we have obtained permits us to draw interesting conclusions:
\begin{itemize}
	\item The trans-series structure of the heat trace on $S^2$, and consequently of the scaling dimensions of large-charge operators of the $O(2N)$ model, is entirely determined by geometric considerations. 
	\item The saddles driving factorial growth in the large-charge expansion do \emph{not} need to be stable. While topological arguments are usually taken as a guideline to the non-perturbative structure of the model, resurgent asymptotics can be driven also by saddles in the same topological class, such as great circles on $S^2$.
     This was already observed for two-dimensional field theories in~\cite{Cherman:2014ofa}. 
	\item The presence of negative modes and the choice of the continuation in Eq.~\eqref{eq:morse} giving rise to the non-trivial Morse index are a geometric realization of the Borel ambiguity in the resurgence computation of Eq.~\eqref{eq:final-heat-trace}.
     Different choices of these phases correspond to the different paths that avoid singularities in the Borel plane in all possible ways.
     This is ultimately related to the Lefschetz thimble decomposition of (path-)integrals~\cite{Witten:2010cx}.
\end{itemize}

\section{Comparison with the small charge expansion}
\label{sec:resurgence}

Up to this point we have discussed the large-\(\hat q\) limit and its resurgence properties.
In fact, in the double-scaling limit, the small-charge regime is directly accessible: one can write a convergent expansion of the grand potential as a function of $\mu$.
Using the definition of the zeta function, the grand potential on the sphere (see Eq.~(\ref{eq:grandPot})) becomes~\cite{Alvarez-Gaume:2019biu}
\begin{multline}
  \omega(\mu) = - \frac{1}{2} \zeta(-\tfrac{1}{2}| S^2, \mu) = - \frac{1}{2} \eval{ \sum_{l=0}^\infty (2l + 1) \left( \frac{l(l+1)}{r^2} + \mu^2 \right)^{-s}}_{s = -1/2} \\
  = -  r^{-2s} \eval{ \sum_{k=0}^\infty \binom{-s}{k} \zeta(2s+2k-1, \tfrac{1}{2}) \pqty{\mu^2 r^2 - \frac{1}{4}}^k}_{s=-1/2},
  \label{eq:pot-small-q}
\end{multline}
where \(\zeta(s,a)\) is the Hurwitz zeta function
\begin{equation}
  \zeta(s,a) = \sum_{n=0}^\infty ( n + a)^{-s}.
\end{equation}
Once again we notice that the correct expansion variable is $m^2 = \mu^2 - \frac{1}{4 r^2 }$.
The formula can be rewritten in terms of Bernoulli numbers:
\begin{equation}
  \omega(m) = r m^2 \sum_{k=0}^\infty (-1)^k  \binom{1/2}{k+1} \frac{(2\pi)^{2k} (2^{2k}-1) B_{2k}}{2 (2k)!} (r m)^{2k}.
\end{equation}
This expansion is convergent rather than asymptotic, with radius of convergence $\abs{rm}  < 1/2$. This corresponds to the value $\mu^2=0$, which turns the $\ell = 0$ mode into a zero mode in the expansion of Eq.~\eqref{eq:pot-small-q}. Importantly, the singularity which determines the radius of convergence is found at the  negative value $m^2 = -1/(4r^2)$. For this reason the expansions for $m^2 \rightarrow 0^+$ and $m^2 \rightarrow \infty$ admit a smooth interpolation for all $m^2 > 0$.

The Legendre relation between $\mu , \hat{q}$ has to be solved order by order.
We give here the first few orders, which can be computed analytically in closed form:
\begin{align}
 r  f(\hat q) &= \frac{\hat q}{2} +\frac{4 \hat q^2}{\pi ^2}+\frac{16 \left(\pi ^2-12\right) \hat q^3}{3 \pi ^4} + \frac{16 \left(384-48 \pi ^2+\pi ^4\right) \hat q^4}{3 \pi ^6} \dots \\
 r \mu &= \frac{1}{2} +\frac{8 \hat q}{\pi ^2} +\frac{16 \left(\pi ^2-12\right) \hat q^2}{\pi ^4} + \frac{64 \left(384-48 \pi ^2+\pi ^4\right) \hat q^3}{3 \pi ^6} + \dots
\end{align}
The free energy is also a convergent expansion, with estimated radius of convergence \(\abs{\hat q} \lessapprox 0.28\dots\). Again, the leading singularity is found at negative values of $\hat{q}$ so that the small and large-charge regimes can be connected without obstructions. This expansion can be in principle computed within perturbation theory around the uncharged vacuum of the $O(2N)$ \ac{cft}.
It turns out to be in general asymptotic, but becomes convergent at leading order in large $N$. 

\begin{figure}
  \centering
  \hfill
  \includegraphics[width=.45\textwidth]{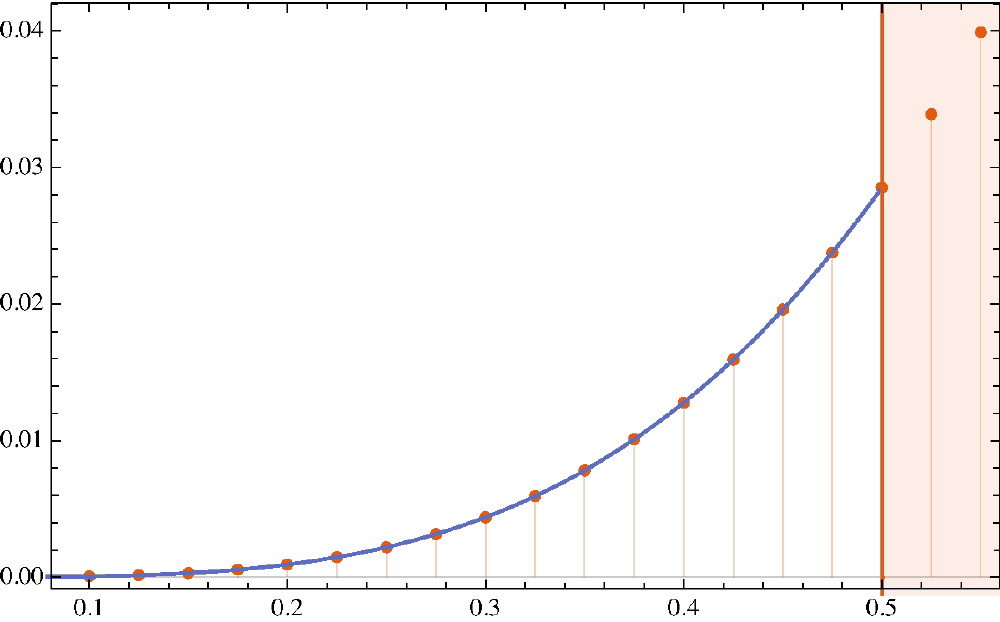}
  \hfill
  \includegraphics[width=.45\textwidth]{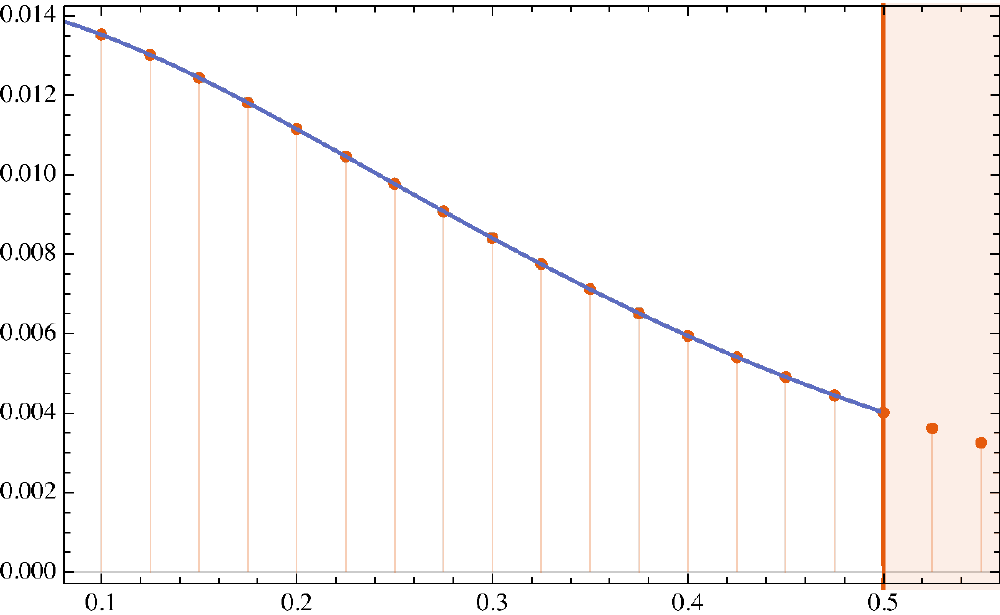}  
  \caption{Real and imaginary part of the lateral Borel summation for the grand potential (dots) compared to the small-charge expansion and the exponential corrections from worldline instantons (continuous line) as function of \(m r \). The two approaches agree completely at the level of resolution of our numerical computation. The small-charge expansion is not convergent in the shaded region.}
  \label{fig:resurgence-vs-convergent}
\end{figure}

This result is to compared with our expression for the grand potential in Eq.~(\ref{eq:resurged-grand-potential}) that we can put into the form
\begin{equation}
  \omega(m) = \operatorname{P.V.}\left[\frac{r m^2}{\pi} {\int_0^\infty \dd{\zeta} \frac{K_{2}(2 m r \zeta)}{\zeta\sin(\zeta)}  }\right].
\end{equation}
In Figure~\ref{fig:resurgence-vs-convergent} we compare the real and imaginary part of the lateral Borel summation of the grand potential to the convergent small-charge expansion and the worldline computation of the exponential correction.
The two approaches agree completely at the level of resolution of our numerical computation.
For example at \(m r = 0.4\), which corresponds to \(\hat q \simeq 0.187\dots \) we find that the small-charge expansion and resurgence agree to at least eight digits:
\begin{align}
  \eval{r \omega(mr=0.4)}_{\text{small charge}} &= 0.012\, 777\,296\,63 \dots \\
  \eval{r \omega(mr=0.4)}_{\text{resurgence}} &= 0.012\,777\,297\,69 \dots
\end{align}
The limiting factor in the resurgence result seems to be computer time: each added digit in precision takes approximately ten times longer than the previous one. 
The relative error is of order $8 \times 10^{-8}$, which is six orders of magnitude smaller than the exponential correction \(e^{-2 \pi \times 0.4} \approx 8 \times 10^{-2}\).
This is strong evidence that we have taken all non-perturbative corrections into account.

\section{Lessons from large N}%
\label{sec:lessons-from-large-N}

In this section we would like to use the exact results that we have obtained in the double-scaling limit~\cite{Alvarez-Gaume:2019biu} to derive some general features of the large-charge expansion.
In the \ac{eft} approach to strongly coupled \acp{cft} at large charge, one studies the system on a manifold \(\mani\) of typical length \(L\) and considers sectors of fixed charge \(Q\).
This allows us to write an \ac{eft} with a cutoff \(\Lambda\) that is bounded by the energy scales \(\Lambda_{\text{low}} = 1/L\) fixed by the geometry, and \(\Lambda_{\text{high}} = Q^{1/d}/L\) fixed by the charge density.
If the hierarchy
\begin{equation}
  \frac{1}{L} \ll \Lambda \ll \frac{Q^{1/d}}{L}    
\end{equation}
is satisfied, the \ac{eft} is weakly coupled and controlled by the ratio of the two scales \(1/Q^{1/d}\).
Lorentz invariance fixes the actual coupling to be \(1/Q^{2/d}\).
Scale invariance fixes the dimension of the terms in the action: for a class of systems that includes the \ac{wf} point, the \ac{eft} is written in terms of a dimensionless field \(\chi\) to be expanded around the ground state at fixed charge \(\chi_0 = \mu t\). The action takes the form
\begin{equation}
  \mathcal{L}_{\ac{eft}} = \omega_0 \pqty{\del_\mu \chi \del^\mu \chi}^{3/2} + \omega_1 \pqty{\del_\mu \chi \del^\mu \chi}^{1/2} + \dots,
  \label{eq:effective-action}
\end{equation}
where we have specialized to the case of \(2 + 1 \) dimensions.
Physical quantities are expressed as series in \(1/Q\).
The energy on the two-sphere for example, which via the state-operator correspondence is proportional to the conformal dimension of the lowest operator of charge \(Q\), takes the form
\begin{equation}
  E = \frac{1}{r} \pqty{ f_0 Q^{3/2} + f_1 Q^{1/2} + \dots },   
\end{equation}
where the coefficients \(f_n\) cannot be computed within the \ac{eft}: they enter as an input and have to be computed independently, \emph{e.g.} in a double-scaling limit~\cite{Orlando:2019hte,Badel:2019oxl,Badel:2019khk,Cuomo:2020rgt,Watanabe:2019pdh,Arias-Tamargo:2019xld,Arias-Tamargo:2019kfr,Arias-Tamargo:2020fow,Antipin:2020abu,Antipin:2020rdw,Antipin:2021akb,Jack:2020wvs,Jack:2021ypd} or on the lattice~\cite{Banerjee:2017fcx,Banerjee:2019jpw}.

In our double-scaling limit, the quantities that we have studied above correspond to the minimal energy configuration for this action. The grand potential is the action evaluated at the minimum:
\begin{equation}
	\omega(\mu) = \eval{S_{\ac{eft}}}_{\chi = \mu t}
\end{equation}
and the free energy is the corresponding ground state energy.
From the formula for the free energy in Eq.~(\ref{eq:sphere-free-energy}) we can infer that in the large-N limit of the \(O(2N)\) \ac{wf} point, the coefficients \(f_n\) are given by
\begin{align}
  f_0 &= \frac{2}{3 (2N)^{1/2}}, & f_1 &= \frac{(2N)^{1/2}}{6} , & f_2 &= -\frac{7 (2N)^{3/2}}{720} _,
\end{align}
and so on.

Even if these coefficients cannot be computed within the \ac{eft}, we can still use our non-perturbative analysis to say something about their generic large-order behavior.
To do so, we make the following assumptions for any \(N\):
\begin{enumerate}
\item the large-charge expansion is asymptotic;
\item the leading singularity in the Borel plane can be obtained as the first non-trivial saddle of a worldline integral for a particle of mass {proportional to} \(\mu\).
\end{enumerate}
This second assumption relies on the fact that we are describing a \ac{cft} which has no intrinsic scales.
The only dimensionful parameter is related to the fixed charge density.%
\footnote{{The proportionality factor is fixed by the mass of the lightest mode in the spectrum. For \(N>1\) one can further conjecture this to be a gapped Goldstone (see~\cite{Cuomo:2020gyl}) whose mass \(\mu\)  is fixed by symmetry and cannot be renormalized by loop effects~\cite{Brauner:2006xm,Nicolis:2012vf}. In this case we would have \(\varpi = 1\) as we have found in the double-scaling limit.}}

This means that we expect the conformal dimensions to take the form of a double expansion in \(1/Q\) and \(e^{-2 \pi {\varpi} r \mu}\):
\begin{equation}
  \label{eq:dimensions-double-series}
  \Delta(Q) = Q^{3/2} \sum_{n} f_n \frac{1}{Q^n} + C_1 Q^{\kappa_1} e^{-3 \pi {\varpi} f_0 \sqrt{Q}} \sum_{n} f^{(1)}_n \frac{1}{Q^{n/2}}  + \dots,
\end{equation}
where \(C_1\),  \(\kappa_1\) and \({\varpi}\) are constants, and we have used the fact that \(\mu\) and \(Q\) are (Legendre) dual variables:
\begin{equation}
  \label{eq:chemical-potential-charge-any-N}
  \mu = \frac{1}{r} \fdv{\Delta(Q)}{Q} = \frac{3}{2 r} f_0 Q^{1/2} + \dots
\end{equation}
Note that the second series is an expansion in \(1/Q^{1/2}\), in agreement with the explicit expression found in Eq.~\eqref{eq:non-perturbative-grand}.\footnote{This trans-series structure, with expansion in $Q^{1/2}$ in the non-perturbative sector, is also found in the Large-$N$ asymptotics in matrix models~\cite{Marino:2008ya,Ahmed:2017lhl}. We thank Gerald Dunne for bringing this fact to our attention.} 

The large-order behavior of the perturbative series in \(1/Q\) is related to the parameters appearing in the non-perturbative part.
For this reason we can turn our conjecture on the form of the exponential term into a universal prediction for the large-order behavior of the \(f_n\).
In general, if the \(f_n\) diverge at large \(n\) as
\begin{equation}
  f_n \sim \frac{(\beta n)!}{A^n} 
\end{equation}
one obtains an optimal truncation of the perturbative series for the value \(N^*\) corresponding to a minimum of \(f_n Q^{-n}\), which in this case is
\begin{equation}
  N^* \approx \frac{1}{\beta} \abs{A Q}^{1/\beta}.
\end{equation}
The error is of order
\begin{equation}
  \epsilon(Q) \sim e^{-(A Q)^{1/\beta}} .
\end{equation}

In our problem we can invert the usual logic.
As we assume the form of the (leading) non-perturbative terms to be the same for any $N$, from Eq.~(\ref{eq:dimensions-double-series}) we read
\begin{align}
  \beta &= 2, & A &= 9 \pi^2 {\varpi}^2 f_0^2,
\end{align}
which preserves the asymptotic growth as $(2n)!$:
\begin{equation}
  f_n \sim \frac{ (2n)!}{(3 \pi {\varpi} f_0)^{n}  } ,
\end{equation}
which is optimally truncated at
\begin{equation}
  N^* \approx \frac{3 \pi {\varpi} f_0}{2} Q^{1/2} .
\end{equation}

The $(2n)!$ divergence is semi-classical  and is the leading contribution to the nonperturbative effects.
The usual instantons associated to quantum corrections grow as $n!$ and are of order $e^{-Q^{3/2}}$, see Appendix~\ref{sec:lipatovs-instantons}.
Note that there is an interplay between the small-\(n\) and large-\(n\) coefficients.
The non-perturbative expansion is associated to the large-\(n\) behavior of the \(f_n\) via resurgence and to the small-\(n\) coefficients via the \ac{eom} in Eq.~(\ref{eq:chemical-potential-charge-any-N}).
This is the reason why we can write the optimal truncation in terms of the lowest coefficient \(f_0\).

\bigskip

This analysis helps to shed some light on the lattice results concerning the \(O(2)\) and \(O(4)\) model~\cite{Banerjee:2017fcx,Banerjee:2019jpw}.
It was remarked in these papers that the large-charge expansion remains very good also for small values of \(Q\) and that a few terms are sufficient to predict the conformal dimensions of the lowest operators.
Generically we expect \(f_0\) to be of order one.
In fact, lattice estimates for the \(O(2)\) and \(O(4)\) model give, respectively, \(f_0 \sim 0.337(3)\) and \(f_0 \sim 0.301(3)\).
Under the previous assumptions, our formula predicts that the optimal truncation is for \(N^* = \order{\sqrt{Q}}\) with an error of order \(\order{e^{-\pi \sqrt{Q}}}\).
This is perfectly consistent with the numerical results.
In~\cite{Banerjee:2017fcx,Banerjee:2019jpw} it has been observed that the first two or three terms in the expansion are sufficient to reproduce the lattice results with great accuracy for charges up to \(Q = \order{10}\).
At \(Q = 1\), the error is of order \(\order{10^{-2}}\) and at \(Q=11\), it is of order \(\order{10^{-5}}\), to be compared to \(e^{-\pi } \approx 4 \times 10^{-2}\) and \(e^{-\pi \sqrt{11}} \approx 3 \times 10^{-5}\).

\section{Conclusions and Outlook}%
\label{sec:conclusions}

In this note, we have used resurgence methods to analyze and extend the large-charge expansion of the O(2N) vector model in three dimensions at the Wilson--Fisher point using results from the double-scaling limit $Q\to \infty$, $N\to \infty$, with $Q/(2N) = \hat q $ constant.
We have studied in detail the cases of the model compactified both on the two-torus and the two-sphere which, via the state-operator correspondence allows to compute the conformal dimension of the lowest operator of charge \(\hat q\).
In both cases we have computed both the perturbation series and the exponentially suppressed non-perturbative corrections.
In the case of the sphere, resurgence leaves us with an ambiguity in the non-perturbative contribution, which we are able to resolve in two ways: via a (simpler) resurgent analysis of the Dawson's function, and using a geometric interpretation in terms of quantum mechanics of particles propagating along geodesics on the compactification manifold.
  The latter method provides an interesting geometric interpretation of the non-perturbative corrections of the problem as well of its Borel ambiguities and   allows us to propose an exact form of the grand potential, valid for any value of \(\hat q\), which we verify to high precision numerically.

The fact that the non-perturbative corrections are finite-volume effects related to the geometry of the compactification manifold seems to be a robust feature independent of the double-scaling limit.
This motivates us further to extend our results to the regime of large charge but finite N, conjecturing that the large-charge expansion is always asymptotic and giving an optimal truncation of { $N^* = \order{\sqrt{Q}}$   
with an error of order \(\epsilon(Q) = \order{e^{- \sqrt{Q}}}\) --- which is} consistent with the lattice results of~\cite{Banerjee:2017fcx,Banerjee:2019jpw}. The nonperturbative contributions that we find are due to the \ac{eft} itself being an asymptotic expansion and dominate over ordinary \ac{qft} instantons.

\medskip
Our observations lead the way to a number of further applications of resurgence in the context of large charge.
The first natural extension is to $d\neq 3$~\cite{Giombi:2020enj}.
Even- and odd-dimensional spheres are known to behave differently~\cite{Camporesi:1990wm} (the heat kernel trace is convergent in odd dimensions) and it would be interesting to see how this reflects in the resurgence analysis at large charge.
Another possibility is to build on the results of a different double-scaling limit, studying the Wilson--Fisher point at
large charge in the $\varepsilon$-expansion~\cite{Badel:2019oxl,Badel:2019khk,Cuomo:2020rgt,Watanabe:2019pdh,Arias-Tamargo:2019xld,Arias-Tamargo:2019kfr,Arias-Tamargo:2020fow,Antipin:2020abu,Antipin:2020rdw,Antipin:2021akb,Jack:2020wvs,Jack:2021ypd}.
In the same spirit, it would be interesting to perform a resurgence analysis of supersymmetric systems, both at large R-charge and in the double-scaling limit $g \to 0$, $Q \to \infty$~\cite{Bourget:2018obm,Grassi:2019txd,Beccaria:2018xxl,Beccaria:2020azj}.

Probably the most interesting issue is the one of the behavior of the large-charge expansion of a generic model. Here we have conjectured a precise relationship between the perturbative \ac{eft} and the leading non-perturbative corrections.
It would be most interesting to prove (or falsify) this and to explore the applicability of this type of reasoning to more general \acp{eft}.

\subsection*{Acknowledgments}

We would like to thank  Luis Álvarez-Gaumé, Daniele Dorigoni, Gerald Dunne, Simeon Hellerman, Marcos Mariño, and Donald Youmans for illuminating discussions and comments on the manuscript.
 
The work of S.R. is supported by the Swiss National Science Foundation under grants number 200021 192137 and PP00P2183718/1.
D.O. acknowledges partial support by the \textsc{nccr 51nf40--141869} ``The Mathematics of Physics'' (Swiss\textsc{map}).

 \appendix

\section{Large charge at large N}%
\label{sec:largeN}

In this appendix, we briefly summarize the results of~\cite{Alvarez-Gaume:2019biu} (for an extended discussion, see the original paper).
Our starting point is the action for the \(O(2N)\) vector model in $2+1$ dimensions at the Wilson--Fisher point,
\begin{equation}
	S[\varphi, \lambda] = \sum_{i=1}^{N} \int \dd{t}\dd{\mani} \left[ \del_\mu\varphi^*_i\del^\mu\varphi_i +(\xi R+\lambda)\varphi^*_i\varphi_i \right],
\end{equation}
where %
$\mani$ is the compact two-dimensional manifold of volume \(V\) on which we are working, $R$ is its Ricci-scalar, $\xi = 1/8$ is the conformal coupling and $\lambda$ is the Lagrange-multiplier that was promoted to a field with a Hubbard--Stratonovich transformation~\cite{stratonovich1957method,Hubbard:1959}.

We want to calculate the canonical partition function for the case that we fixed the charge corresponding to the Cartan generator which rotates the field $\varphi_N$, which corresponds to restricting our attention to the completely symmetric representation of rank \(Q\) (for an extended discussion of charge fixing in the O(2N) vector model, see Section~4.1 in~\cite{Gaume:2020bmp}, see also~\cite{Antipin:2020abu}).
We have
\begin{equation}
	Z(Q) = \Tr\left( e^{-\frac{1}{T} H}\delta(Q-\hat Q)\right) %
	= \int \frac{\dd{\theta}}{2\pi} e^{i\theta Q} \DD{\varphi_i}\DD{\varphi^*_i}\DD{\lambda} e^{-S_\theta[\varphi,\lambda]},
\end{equation}
where
\begin{align}
	S_\theta[\varphi,\lambda] = \int \dd{t}\dd{\mani} \left[\sum_{i=1}^{N-1} \del_\mu\varphi^*_i\del^\mu\varphi_i +(D_\mu\varphi_N)^*D^\mu\varphi_N + (\xi R+\lambda)\sum_{i=1}^{N}\varphi^*_i \varphi_i\right]
\end{align}
with 
\begin{equation}
	D_\mu \varphi = \begin{cases}
		(\del_0+iT\theta)\varphi\\
		\del_i\varphi,
	\end{cases}
\end{equation}
where $T$ is the inverse length of the time circle.
The integral of the first $N-1$ fields $\varphi_i$ is quadratic and can be performed to obtain an effective action for the fields \(\varphi_N\) and \(\lambda\):
\begin{equation}
  Z(Q) = \int \frac{\dd{\theta}}{2\pi} e^{i\theta Q} \DD{\varphi_N} \DD{\varphi_N^*} \DD{\lambda} e^{-S_\theta[\varphi_N,\lambda]},
\end{equation}
where
\begin{multline}
  S[\varphi_N,\lambda] = (N-1) \Tr[\log (-\del_0^2{} - \Laplacian{} + \xi R+ \lambda) ] +\\
  +\int \dd{t} \dd{\mani} \left[(D_\mu\varphi_N)^*D^\mu\varphi_N + (\xi R+\lambda)\varphi^*_N \varphi_N \right].
\end{multline}
The path integral localizes around the saddle point obtained from minimizing the action w.r.t. $\theta$ and the zero modes of $\varphi_N$ and $\lambda$.
We expand the fields into vev and fluctuations:
\begin{align}
	\varphi_N &= \frac{A}{\sqrt{2}} + u, & \lambda&= \mu^2 - \xi R +\hat \lambda = m^2 + \hat \lambda.
\end{align}
The parameter \(\mu\) is the mass with respect to the Laplace--Beltrami operator \(\Laplacian{}\), while \(m\) is the mass with respect to the conformal Laplacian \(\Laplacian{} - \xi R\).

The free energy is obtained by minimizing
\begin{multline}\label{eq:SQ}
	S_Q = -i\theta Q + (N-1)\Tr[\log (-\del_0^2{}-\Laplacian{} + \mu^2 + \hat \lambda) ] +\\
	+\int \dd{t} \dd{\mani} \left[(D_\mu u)^*D^\mu u + \frac{A^2\theta^2T^2}{2} + (\mu^2+\hat\lambda)\left|\frac{A}{\sqrt{2}}+u \right|^2\right],
\end{multline}
setting the fluctuations to zero:
\begin{equation}
  S_Q^{\text{saddle}} = -i\theta Q +(N-1)\Tr[\log (-\del_0^2{}- \Laplacian{} + \mu^2) ] + \frac{1}{T} V \frac{A^2}{2}\pqty{\theta^2T^2 + \mu^2}.
\end{equation}
This last expression needs to be minimized:
\begin{equation}
	\begin{cases}
		\del_\theta: & -iQ+VA^2\theta T= 0,\\
		\del_\mu: & (N-1)\pdv{\mu} \Tr[\log (-\del_0^2{}-\Laplacian{} +  \mu^2) ] + \frac{1}{T} V A^2\mu =0,\\
		\del_A: & {\theta^2T^2} + \mu^2 = 0,
	\end{cases}
\end{equation}
which we can rearrange as
\begin{equation}
	\begin{cases}
		\theta T= i \mu,\\
		Q = -i {V}A^2 \theta T = - VA^2\mu = (N-1)\del_\mu \Tr\left[\log (-\del_0^2{} - \Laplacian{}+ \mu^2) \right].
	\end{cases}
\end{equation}
If we take the double scaling limit
\begin{align}
	Q &\to \infty, & N&\to \infty, & \frac{Q}{2N}=\hat q \ \text{fixed}, 
\end{align}
the first line of Eq.~\eqref{eq:SQ} dominates and the path integral localizes around the saddle.

The free energy at the saddle takes the form
\begin{equation}
	F(\hat q) = -T\log(Z(Q)) = 2N\left[ \mu\hat q + \frac{T}{2}\log(\det(-\del_0^2{}-\Laplacian{}+\mu^2)) \right] +\order{N^0},
\end{equation}
where $\mu^2$ is to be understood as a function of $\hat q$ via
\begin{equation}
  \hat q = \pdv{\mu} \left( \frac{T}{2}\log(\det(-\del_0^2{}-\Laplacian{}+\mu^2)) \right) .
\end{equation}
One can read these equations as the free energy being the Legendre transform of the functional determinant, which is naturally identified with the grand potential (Landau free energy)\footnote{Since we are working at leading order in \(N\) it is convenient to use the free energy and grand potential per degree of freedom. The corresponding extensive quantities are related by a similar Legendre transformation \(F(Q) = \sup_\mu ( \mu Q - \Omega(\mu))\).}
\begin{equation}
  f(\hat q) = \frac{F(\hat q)}{2N} =\sup_{\mu }(\mu \hat q - \omega(\mu)),
\end{equation}
where
\begin{equation}
  \omega(\mu) = - \frac{T}{2}\log(\det(-\del_0^2{}-\Laplacian{}+\mu^2)) .
\end{equation}

We are interested in studying the theory on different manifolds, so it is convenient to write the functional determinant in terms of zeta functions%
\begin{equation}
  \log(\det(- \del_0^2{} - \Laplacian{} + \mu^2 )) = - \eval{ \dv{s} \zeta(s | S^1 \times \mani, \mu)}_{s=0},
\end{equation}
where \(\zeta(s| S^1 \times \mani, \mu)\) is the zeta function that we can write as a Mellin integral:
\begin{equation}
  \label{eq:2}
  \zeta(s | M, \mu) = \frac{1}{\Gamma(s)}  \int_0^\infty \frac{\dd{t}}{t} t^s \Tr(e^{(\del_0^2{} + \Laplacian{} - \mu^2) t} )
\end{equation}
so that
\begin{equation}
  \label{eq:Schwinger-representation}
     \log(\det(- \del_0^2{} - \Laplacian{} + \mu^2 )) = -  \int_0^\infty \frac{\dd{t}}{t}  \Tr(e^{(\del_0^2{} + \Laplacian{} - \mu^2) t} ).
\end{equation}
Using the fact that our geometry is a product \(S_{1/T}^1 \times \mani\), we can separate the \(S^1\) part.
The corresponding heat kernel trace is a theta function:
\begin{equation}
  \Tr( e^{\del_0^2{} t}) = \sum_{n \in \setZ} e^{-{4 \pi^2 n^2 T^2} t} = \theta_3(0, e^{-{4 \pi^2 T^2 t}}) = \frac{1}{T\sqrt{4 \pi  t}} \pqty{1 +  \sideset{}{'}\sum_{k \in \setZ} e^{- \frac{k^2}{4 T^2 t} } } ,
\end{equation}
where the prime indicates that the sum runs over the non-zero modes.
At zero temperature \(T \to 0\) the sum can be approximated with
\begin{equation}
  \Tr( e^{\del_0^2{} t}) \sim \frac{1}{T\sqrt{4 \pi t}}. 
\end{equation}
Then we can write the functional determinant in terms of the Laplace transform of the heat kernel on \(\mani\):
\begin{equation}
  \log(\det(- \del_0^2{} - \Laplacian{} + \mu^2 )) = - \frac{1}{T\sqrt{4 \pi}} \int_0^\infty \frac{\dd{t}}{t} t^{-1/2} e^{-\mu^2 t} \Tr( e^{\Laplacian{} t}) = \frac{1}{T} \zeta(-\tfrac{1}{2} | \mani, \mu).
\end{equation}
The grand potential is
\begin{equation}
  \label{eq:grand-potential-zeta}
  \omega(\mu) = -\frac{1}{2} \zeta(-\tfrac{1}{2} | \mani, \mu),
\end{equation}
and the free energy is
\begin{equation}
  F(\hat q) = 2N \bqty{ \mu \hat q +\frac{1}{2}  \zeta(-\tfrac{1}{2} | \mani, \mu) }.  
\end{equation}
\(\mu\) is the solution to
\begin{equation}
  \hat q  = - \mu \zeta(\tfrac{1}{2}| \mani, \mu),
\end{equation}
where we have used the difference-differential equation
\begin{equation}
  \dv{\mu} \zeta(s| \mani, \mu) = - 2 \mu \zeta(s + 1 | \mani, \mu) .
\end{equation}
These expressions are exact in the limit of \(N \to \infty\) for any value of $\hat q = Q/N$ and can be interpreted as a semiclassical resummation of infinite \(1/N\) corrections.

\section{The Borel transform}
\label{sec:borel-transform}

The \emph{Borel transform} is an operation acting in the space of power series as follows:
\begin{align}
\Phi(z) &\sim \sum_{n=0}^{\infty} a_n z^n &\longrightarrow &&  \mathcal{B}\left\{\Phi\right\}(\zeta) &= \sum_{n=0}^{\infty} \frac{a_n}{\Gamma(\beta n + b)} \zeta^n,
\label{eq:general-Borel}
\end{align}
where we assumed a generic large order behavior of $a_k$ as in Eq.~\eqref{eq:GenericLargeOrder} with $b \equiv \max\{ b_k\},\, \beta \equiv \max \{ \beta_k\}$. This ensures that the series defined by $\mathcal{B}\left\{ \Phi \right\}$ is convergent in a disc centered at the origin of the $\zeta$-plane, also denoted as \emph{Borel plane}.
The analytic properties of the Borel transforms can be directly inferred from Eq.~\eqref{eq:GenericLargeOrder}.
For example, if we have
\begin{align}
	\frac{a_n}{\Gamma(\beta n + b)} &\sim \frac{1}{A^{\beta n +b}} & \text{then} && \mathcal{B}\{ \Phi\} (\zeta ) &\xrightarrow{\zeta \rightarrow A^\beta} \frac{A^{-b}}{1-\zeta\, / A^{\beta}} + \text{regular}.
	\label{eq:small-example}
\end{align}
It is then possible to define the \emph{Borel resummation} of $\Phi$ as 
\begin{equation}
\mathcal{S}\{ \Phi\} (z) = \frac{1}{ \beta} \int_0^\infty \frac{\dd{\zeta}}{\zeta}  \left( \frac{\zeta}{z} \right)^{\frac{b}{\beta}} e^{- \left(\zeta/z \right)^{1/\beta}} \mathcal{B}\{\Phi\}(\zeta).
\label{eq:def-Borel}
\end{equation}
Using the definition of Gamma function, it is evident that $\mathcal{S}\left\{ \Phi \right\}(z) \sim \Phi(z)$ as $z \rightarrow 0^+$. However, if the integral %
is well-defined, it defines a function computable for all values of $z$, which represents the ``sum'' of the divergent series $\Phi$.
This function is unambiguous unless $\mathcal{B}\{\Phi\}$ presents singularities along the integration path.\footnote{The behavior at $\zeta \rightarrow 0, \infty$ is relevant as well, but always regular enough for the cases we study.} In this case one needs to define a directional summation $\mathcal{S}_\theta$ by integrating along the ray with angle $\theta$ in the Borel plane: the formula in Eq.~\eqref{eq:def-Borel} corresponds to the case $\theta= 0$. 

A ray $\theta$ along which $\mathcal{B}\{\Phi\}$ has singularities is a \emph{Stokes ray} and the Borel resummation becomes ambiguous there.
One then defines lateral summations $\mathcal{S}_{\theta^{\pm}}$ by deforming the contour of integration to avoid the singularities. This indicates the emergence of a branch cut for $\operatorname{Arg}(z) = \theta$ for the Borel resummation, with a discontinuity computed as
\begin{equation}
[\mathcal{S}_{\theta^+} - \mathcal{S}_{\theta^-} ]\{\Phi\} = - [ \mathcal{S}_{\theta^-} \circ \Disc_\theta ] \{ \Phi\}.
\end{equation}
This discontinuity is purely non-perturbative, as one can see from the example in Eq.~\eqref{eq:small-example} where $\theta = 0$ \footnote{Notice that $\mathcal{S}$ acts formally as an operator on power series with integer positive powers only, so that here we use $\mathcal{S}_{0^-} \{ 1 \} = 1$} :
\begin{equation}
[\mathcal{S}_{0^+} - \mathcal{S}_{0^-} ]\{\Phi\}  = - \frac{2\pi i }{\beta } z^{- b/\beta} e^{-A/z^{1/\beta}} = - \Disc_0 \{ \Phi\}.
\end{equation}
Given a non-Borel summable perturbative series $\Phi^{(0)}$, the quantity $\Disc\{ \Phi^{(0)}\} $ provides the structure of non-perturbative terms that one has to add to the trans-series completion in Eq.~\eqref{eq:trans-series}, and the large-order behavior in Eq.~\eqref{eq:GenericLargeOrder} follows applying Cauchy's integral representation.  
The lateral Borel summation of the general trans-series is defined as
\begin{equation}
\mathcal{S}_{\theta^\pm} \{	\Phi \} (\sigma_k,z) = \mathcal{S}_{\theta^\pm} \{ \Phi^{(0)} \} (z) + \sum_{k} \sigma_k^\pm e^{- A_k /z^{1/\beta_k}} z^{-b_k / \beta_k} \, \mathcal{S}_{\theta^\pm} \{\Phi^{(k)} \} (z).
	\label{eq:Borel-trans-series}
\end{equation} 
This will not define a unique resurgent function for $\text{Arg}(z) = \theta$ unless we somehow fix the (non-perturbative) ambiguity related to the integration path.
This can be done by imposing extra conditions on the lateral Borel sums,  which determine the trans-series parameters $\sigma_k^\pm$ in such a way that the ambiguity is removed. When this is possible, we achieved ``semiclassical decoding'' in the terminology of~\cite{Marino:2021six}.

\section{Lipatov's instantons}
\label{sec:lipatovs-instantons}

The non-perturbative effects that we have discussed in this work come from the fact that the effective action is by itself an asymptotic expansion, meaning that its Wilson coefficients are responsible for the factorial growth.
This is different from the ordinary factorial growth related to the presence of instantons, and manifest themselves in the diagram proliferation in the quantum perturbative expansion.
Here we follow the discussion of Lipatov~\cite{Lipatov:1976ny} to show that these effects are subleading with respect to the ones we have discussed.

Consider the leading term in the effective action in Eq.~\eqref{eq:effective-action},
\begin{equation}
  \mathcal{L} = \omega_0 \norm{\dd{\chi}}^3.
\end{equation}
To quantize we expand around the fixed-charge solution \(\chi = \mu t\).
It is convenient to expand \(\chi \) into ground state and massless fluctuations \(\hat \chi\), rescaled as
\begin{equation}
  \chi(t, x) = \mu t + \mu \hat \chi(t,x).
\end{equation}
The action is then
\begin{equation}
  \mathcal{L}(\chi) = \omega_0 \mu^3 \left[ ( 1 + \del_0 \hat \chi)^2 - (\del_i \hat \chi)^2 \right]^{3/2} \equiv \omega_0 \mu^3 \mathcal{L}(\hat \chi).
\end{equation}
Perturbation theory around the solution $\hat{\chi} = 0$ is defined by expanding the Lagrangian in power series of derivatives.
The loop counting parameter is \(g \equiv 1/(\omega_0\mu^3)\) since propagators are proportional to \(1/g\) and any vertex carries a factor of \(g\).
The (grand-canonical) partition function is a series in \(g\):
\begin{equation}
  Z(g) = \int \DD{\hat \chi} e^{- \frac{1}{g} S[\hat \chi]} = \sum_{n=0}^\infty Z_{n} g^n.
\end{equation}
We can estimate the leading behavior of the \(Z_n\) for large \(n\) the expression in terms of a contour integral around \( g = 0\):
\begin{equation}
  Z_n = \frac{1}{2 \pi i} \oint \frac{\dd{g}}{g} \frac{Z(g)}{g^n}. 
\end{equation}
For \(n \to \infty\) the integral can be solved via saddle point approximation for both the path integral and the contour integral:
\begin{equation}
  Z_n = \frac{1}{2 \pi i} \oint \frac{\dd{g}}{g} \int \DD{\hat \chi} \exp[- \frac{1}{g} S[\hat \chi] - n \log(g)].
\end{equation}
The saddle point equations are
\begin{equation}
  \begin{cases}
    \fdv{S[\hat \chi]}{\hat \chi} = 0, \\
    \frac{1}{g^2} S[\hat \chi] - \frac{n}{g} = 0 . 
  \end{cases}
\end{equation}
Assuming the existence of a finite-action solution to these equations, the value \(\bar S\) of \(S[\hat \chi]\) at the saddle does not depend on \(n\) or on \(g\), so the second equation implies that at the saddle  \(g \sim 1/n \). Then one can see that
\begin{equation}
  \begin{aligned}
    Z_n g^n & \approx \eval{\exp[- \frac{1}{g} S[\hat \chi ] - n \log(g) ]}_{\text{saddle}} g^n = e^{n \log(n) - n(\log(\bar S) +  1)}  \approx \frac{n!}{\sqrt{2 \pi n} \bar S^n} g^n \\
    &\approx n! \mu^{-3n} \approx n! Q^{-3n/2}.
  \end{aligned}
\end{equation}
We find that the quantum effects lead to a divergent series in \(g\) whose coefficients grow like \(n!\) (as opposed to the \((2n)!\) that we have encountered in Section~\ref{sec:perturbative-sphere}).
This is the growth related to diagram proliferation when one takes the leading order of Eq.~\eqref{eq:effective-action} as an action. 
By the usual resurge arguments, this corresponds to a non-perturbative effect of order
\begin{equation}
  \omega(\mu) \approx \order{e^{-  \mu^3}} \approx \order{e^{-Q^{3/2}}}
\end{equation}
which is parametrically smaller than the \(\order{e^{-Q^{1/2}}}\) semiclassical effects.

\section{Optimal truncation in the double-scaling limit}
\label{sec:optimal-truncation}

In the double-scaling limit we know all the series coefficients of $\omega$, so it is possible to compute the Borel transform and evaluate the free energy to any given precision. This is not possible in the generic \ac{eft} arising at finite $N$, where we need to resort to numerical calculations on the lattice to find the first few coefficients. In this appendix, we show what kind of results to expect for the optimal truncation using the results in the double-scaling limit. More sophisticated analysis such as Borel-Pad\'e and conformal mappings~\cite{Costin:2019xql} can lead to significant improvement on the optimal truncation estimates.

\begin{figure}
  \centering
  \includegraphics[width=.48\textwidth]{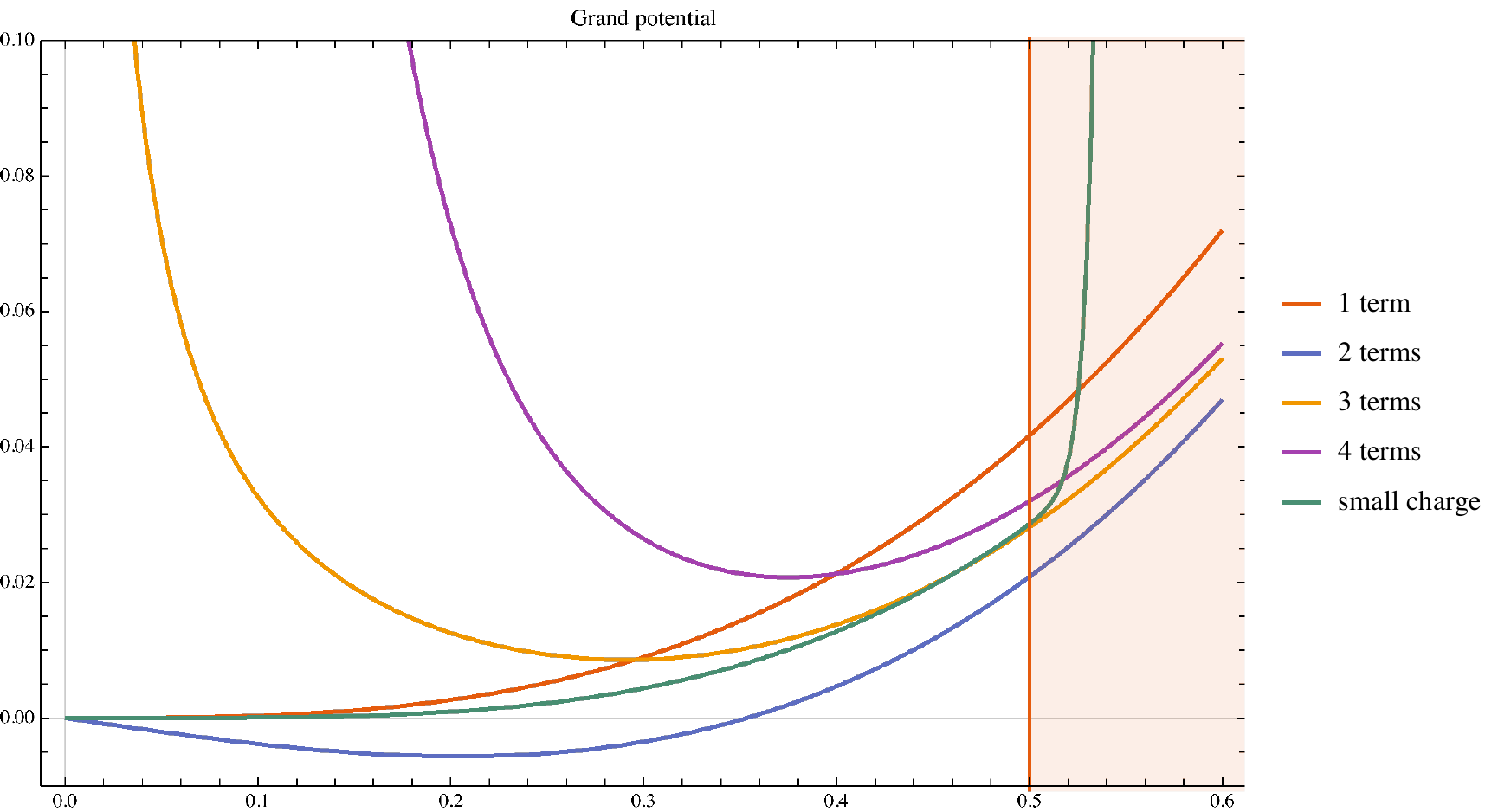} \hfill
  \includegraphics[width=.48\textwidth]{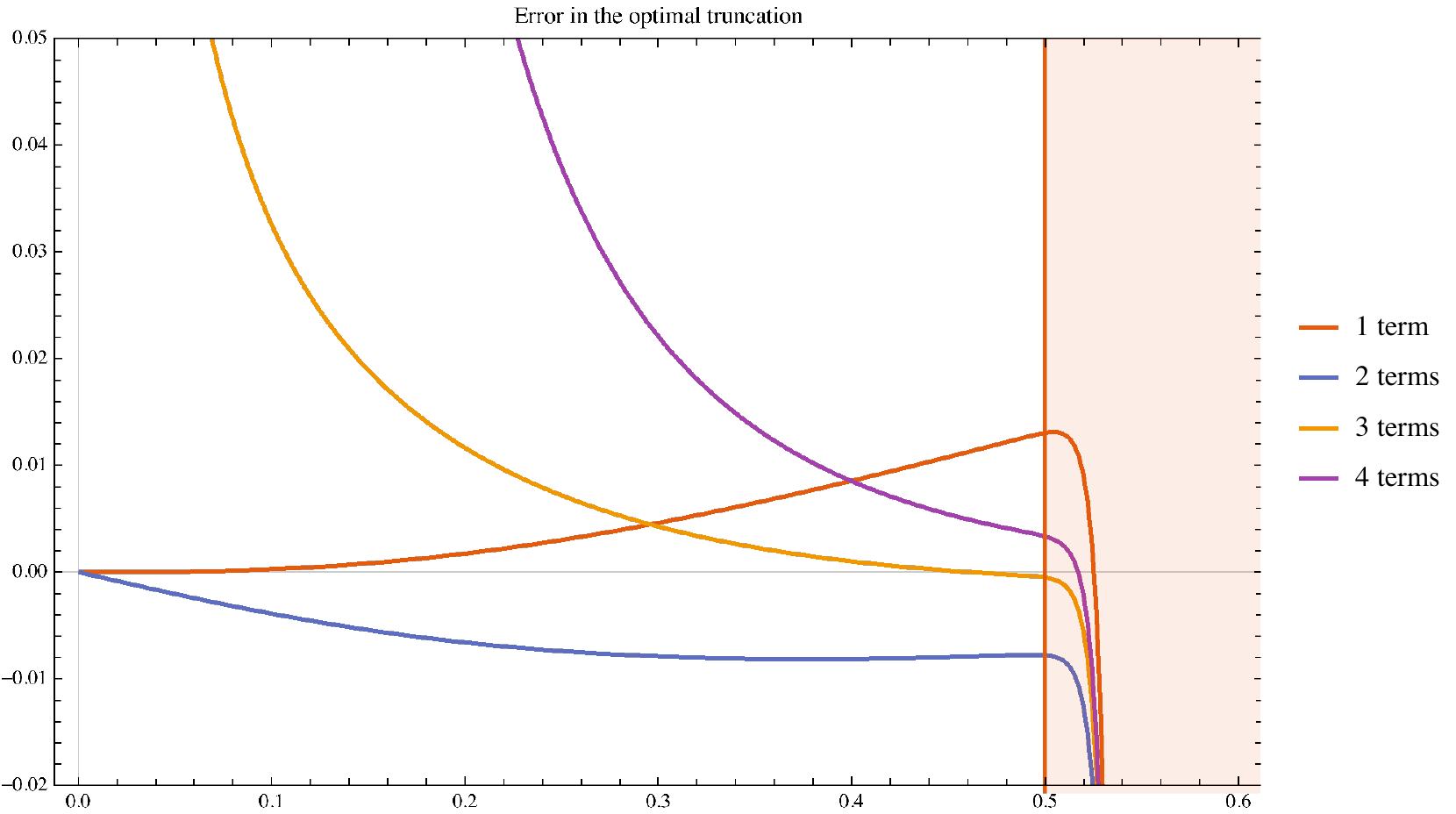}
  \caption{Left: Grand potential $\omega$ as function of $m$ in the small charge expansion and for different numbers of terms. The small-charge expansion breaks down in the red-shaded region. Right: Error in the truncation w.r.t. the exact small-charge expansion.}
  \label{fig:asymptotic-vs-convergent}
\end{figure}

Consider the grand potential \(\omega(m)\) in Eq.~(\ref{eq:large-m-grand-potential}):
\begin{equation}
  \omega(m) = r^2 m^3 \sum_{n=0}^\infty \frac{(-1)^n \pqty{1 - 2^{1 - 2n}} B_{2n} \Gamma(\frac{3}{2}-n )}{2 \sqrt{\pi} \Gamma(n+1)} \frac{1}{(rm)^{2n}} =  r^2 m^3  \sum_{n=0}^\infty \omega_n \frac{1}{(rm)^{2n}}.
\end{equation}
In the main text we show that if for large values of \(n\) the coefficients \(\omega_n\) grow like
\begin{equation}
  \omega_n \sim (\beta n)! A^{-n},
\end{equation}
then the series has an optimal truncation given by the value of \(n\) for which \(\omega_n (rm)^{-2n}\) has a saddle:
\begin{equation}
  \label{eq:optimal-truncation}
  N^* \approx \frac{1}{\beta} \abs{A (rm)^2}^{1/\beta},
\end{equation}
and the error in the truncation is of the order 
\begin{equation}
  \epsilon(m) \sim e^{-(A r^2 m^2)^{1/\beta}} .
\end{equation}
In our case, for large \(n\) we have
\begin{equation}
  \omega_n \sim \frac{ (2n )! (4 \pi^2)^{-n} }{\sqrt{\pi} n^{5/2}} ,
\end{equation}
so \(A = 4 \pi^2\), \(\beta = 2\) and the optimal truncation is for
\begin{equation}
  N^* \approx \pi r m .
\end{equation}
Using the fact that for large values of  \(\hat q\), the coefficient \(m\) scales as \(r m \sim \sqrt{\hat q} \ \) (see Eq.~(\ref{eq:Legendre})), we find that in the expansion of grand potential, and  of the free energy, the optimal truncation is found at
\begin{equation}
  N^* \approx \pi \sqrt{\hat q} . 
\end{equation}
In Figure~\ref{fig:asymptotic-vs-convergent} we compare the asymptotic expansion truncated at the \(N\)-th term with the convergent small-charge expansion.
Empirically we observe that the optimal truncation for \(m < 0.5\) (corresponding to \(\hat q \simeq 0.314\dots\)) is at \(N = 3\) terms until we hit the boundary where the convergent expansion cannot be trusted anymore.
This is quite close to the asymptotic estimate above that would give \(N^* \lessapprox 3\) in this regime.

\section{Trans-series representation of the Dawson's function}
\label{sec:Dawson}

In this appendix we construct the trans-series representation of the Dawson's function appearing in Section~\ref{sec:perturbative-sphere}.
We will show that no non-perturbative ambiguities are left to fix once we impose the reality condition on the heat trace of $S^2$.
The treatment is standard and follows the one used in Euler's and Riccati's \ac{ode}~\cite{Dorigoni:2014hea}.
The Dawson's function is the unique solution of the following Cauchy problem:
\begin{align}
	\dv{F}{z} + 2z F(z) &= 1, & F(0) &= 0.
\end{align}
It is evident that $z = \infty$ is a critical point of the \ac{ode}.
One may attempt to find a solution in terms of an asymptotic series around this point, which turns out to be
\begin{align}
F(z) &\sim \sum_{k=0}^\infty a_n  \frac{1}{z^{2k+1}} =  \sum_{k=0}^\infty \frac{(2k-1)!!}{2^{k+1}} \frac{1}{z^{2k+1}} & \text{for } z &\rightarrow \infty.
\label{eq:Dawson-asymptotic}
\end{align}
This solution has the following problems: (a) it is an asymptotic series, and it holds only at the formal level; (b) there is no constant of integration one can fix via the initial condition.

These problems have a common solution. Define the following Borel transform;
\begin{equation}
	\mathcal{B}\{ \Phi\}(\zeta) =  \sum_{k=0}^\infty \frac{a_n}{\Gamma \left( k + \frac{1}{2} \right)} \zeta^{k+1}=\frac{1}{2\sqrt{\pi}} \frac{\zeta}{1-\zeta},
\end{equation}
The series is not Borel summable because of the singularity at $\zeta =1$, and the appropriate lateral Borel summations are
\begin{equation}
	\mathcal{S}_{\pm} \{ \Phi\} (z) = \int_{\pm} \dd{\zeta} \zeta^{-\frac{3}{2}} e^{-z^2 \zeta} \mathcal{B}\{ \Phi\}(\zeta).
\end{equation}
The discontinuity between the lateral summations reads
\begin{equation}
\left[\mathcal{S}_{+} - \mathcal{S}_-\right]\{ \Phi\}(z)   =  - \Disc \{ \Phi \}(z) = - \frac{2\pi i}{2\sqrt{\pi}} e^{-z^2}.
\end{equation}
This expression is a solution to the homogeneous \ac{ode}.
We can then write the general solution to the \ac{ode} as a trans-series centered at $z = \infty$ as with a single trans-series parameter $\sigma$ as follows:
\begin{equation}
	\Xi (z,\sigma) =  \sum_{k=0}^\infty \frac{(2k+1)!!}{2^{k+1}} \frac{1}{z^{2k+1}} + \sigma e^{-z^2}.
\end{equation}
Its lateral Borel sum reads
\begin{equation}
	\mathcal{S}_{\pm} \left\{ \Xi \right\} (z,\sigma_{\pm}) = \frac{1}{2\sqrt{\pi}} \int_{\pm} \frac{\dd{\zeta}}{\sqrt{\zeta}} \frac{e^{-z^2 \zeta}}{1-\zeta} + \sigma_{\pm } e^{-z^2},
\end{equation}
which is easily proven to be a solution of the inhomogeneous \ac{ode}.

To fix the trans-series parameter we can impose the result to be real for $z$ real, which sets 
\begin{equation}
	\Im \sigma_{\pm} =\pm \frac{ \pi}{2 \sqrt{\pi}},
\end{equation}
so that the coefficients of the non-perturbative exponentials are purely imaginary. 
The real part of $\sigma_{\pm}$ is left unfixed and corresponds to the integration constant which had disappeared when we had attempted a power series ansatz for the solution.
A general boundary condition sets its value as
\begin{align}
	F(0) &= c, & \implies && \Re \sigma_{\mp} &= \frac{\sqrt{\pi}}{2} e^{-c^2} \operatorname{erfi}(c).
\end{align}
For $F(0) = 0$ we have $\Re \sigma_{\pm} = 0$, so that the reality condition is sufficient to fix the whole non-perturbative ambiguity. 
Also note that this fixed version of the trans-series resummation coincides with the definition of the Dawson's function:
\begin{equation}
	\mathcal{S}_{\pm} \left\{ \Xi \right\} (z,\sigma_{\pm}) = \frac{1}{2 \sqrt{\pi}} \operatorname{P.V.} \int_{0}^\infty \dd{\zeta} \frac{e^{-z^2 \zeta}}{\sqrt{\zeta}(1-\zeta)} = \frac{\sqrt{\pi}}{2}e^{-z^2} \operatorname{erfi}(z) \equiv F(z), \quad z >0.
\end{equation}
Since the factorial growth of the Dawson's function is the one driving the one observed in the heat trace, grand potential and free energy, we can conclude that a reality condition is sufficient to fix all non-perturbative ambiguities.

\newpage
\setstretch{1}
\printbibliography{}

\end{document}